\documentclass[]{emulateapj}

\usepackage{amsmath}

\newcommand{\kms}{\ensuremath{\mathrm{km~s}^{-1}}}
\newcommand{\Nifs}{\ensuremath{^{56}\mathrm{Ni}}}

\newcommand{\texp}{\ensuremath{t_{\mathrm{ej}}}}

\newcommand{\msun}{\ensuremath{M_\odot}}

\newcommand{\dls}{\ensuremath{\Delta \lambda_i}}
\newcommand{\dvs}{\ensuremath{\Delta v_s}}
\newcommand{\aex}{\ensuremath{\alpha_{\rm ex}}}
\newcommand{\fosc}{\ensuremath{f_{\rm osc}}}

\newcommand{\csg}{\ensuremath{{\rm cm^2~g^{-1}}}}
\newcommand{\gcc}{\ensuremath{{\rm g~cm^{-3}}}}

\newcommand{\AS}{\texttt{Autostructure}}
\newcommand{\Sedona}{\texttt{Sedona}}

\begin{document}

\title {Opacities and Spectra of the  r-process Ejecta from Neutron Star Mergers }
\shorttitle{r-process opacities}
\shortauthors{Kasen, Badnell, and Barnes}

\author{Daniel Kasen$^{1,2}$, N. R. Badnell$^3$, Jennifer Barnes$^{1,2}$} 
\affil{$^1$  Department of Physics and Astronomy,
University of California, Berkeley, CA 94720, USA}
\affil{$^2$ Nuclear Science Division, Lawrence Berkeley National Laboratory, 1 Cyclotron Road, Berkeley, CA 94720, USA}
\affil{$^3$ Department of Physics, University of Strathclyde, Glasgow G4 0NG, UK}

\begin{abstract}
Material ejected during (or immediately following) the merger of two
neutron stars may assemble into heavy elements by the r-process.  The
subsequent radioactive decay of the nuclei can power 
electromagnetic emission similar to, but significantly dimmer than, an
ordinary supernova.  Identifying such events is an important goal of
future transient surveys, offering new perspectives on the origin of
r-process nuclei and the astrophysical sources of gravitational waves.
Predictions of the transient light curves and spectra, however, have
suffered from the uncertain optical properties of heavy ions.  Here we
consider the opacity of expanding r-process material 
and argue that it is dominated
by line transitions from those ions with the most complex
valence electron structure, namely the lanthanides.  For a few
representative ions, we run atomic structure models to calculate
radiative data for tens of millions of lines.  We find that the
resulting r-process opacities are orders of magnitude larger than that
of ordinary (e.g., iron-rich) supernova ejecta.  Radiative transport
calculations using these new opacities indicate that the transient emission
should be dimmer and redder than previously thought.  The
spectra appear pseudo-blackbody, with broad absorption features, and
peak in the infrared ($\sim 1~\mu$m). We discuss uncertainties in the
opacities and attempt to quantify their impact on the spectral
predictions.  The results have important implications for
observational strategies to find and study the radioactively powered
electromagnetic counterparts to compact object mergers.

\end{abstract}

\section{Introduction}
\label{sec:intro}

Hydrodynamical simulations suggest that a small fraction of mass is
ejected when two neutron stars (or a black hole and neutron star)
collide or merge \citep{Janka_1999,Rosswog_1999,Lee_2001,
  Rosswog_2005,Oechslin_2007, Chawla_2010,Shibata_2011, Hoto_2013}.
If this ejecta is sufficiently neutron rich, it will assemble within
seconds into heavy elements via rapid neutron captures (the r-process)
\citep{Lattimer_1974,Eichler_1989, Freiburghaus_1999}.  The subsequent
beta decay of the nuclei will heat the ejecta for days, powering a
thermal, supernova-like transient \citep{Li_Paz_1998}.  Because the
ejected mass is small in comparison to ordinary supernovae (SNe), the
light curves of these ``r-process supernovae" are expected to be
relatively dim and short lived.  Previous radiative models predict
peak bolometric luminosities around $10^{40}-10^{42}~{\rm
  ergs~s^{-1}}$, peaking at optical wavelengths and lasting around a
day
\citep{Li_Paz_1998,Kulkarni_2005,Metzger_2010,Roberts_2011,Goriely_2011,
  Piran_2012}.

Although we have not yet discovered an r-process SN from a neutron
star merger (NSM), there are compelling reasons to look for them.
Because these outflows are non-relativistic, they emit radiation
relatively isotropically, and are therefore promising electromagnetic
counterparts to gravitational wave sources; if discovered
coincidently, they could enhance the scientific value of an advanced
LIGO/VIRGO gravitational wave signal
\citep{Schutz_1986,Kochanek_1993,Sylvester_2003,Phinney_2009,Mandel_2010,Metzger_Berger_2011,
  Kelley_2012,Nissanke_2012}.  Discovery of r-process SNe would also
dramatically illuminate our incomplete understanding of heavy element
production in the Universe.  The NSM ejecta is thought to be a
remarkably pure sample of r-process material, which would allow us to
cleanly study heavy elements near their production site, and soon
after they had been created.  In principle, analysis of the light
curves and spectra of these radioactive transients could be used to
quantify the mass and chemical composition of the ejecta, which would
clarify the unknown site(s) of r-process nucleosynthesis
\citep[e.g.,][]{Arnould_2007,Sneden_2008}.

Perhaps the largest remaining uncertainty in our understanding of r-process SNe
has concerned the opacity of the ejected debris, which (along with the
ejecta mass and kinetic energy) is a key parameter determining the
brightness, duration, and color of the transient.  The ejecta of NSMs
consists of heavy elements in rapid differential expansion, and at
relatively low densities and temperatures ($\rho \sim 10^{-13}~\gcc$
and $T \sim 5000$~K at 1 day after ejection).  Because almost nothing
is known about the optical properties of such material, previous
radiative transfer models have simply adopted opacities characteristic
of ordinary SNe.  In Type~Ia supernovae (SNe~Ia), for example, the
opacity is due primarily to numerous iron group lines, which are
blended by Doppler broadening into a pseudo-continuum.  We can expect
that lines will also dominate the opacity of NSM ejecta, but
unfortunately very little atomic data exists for ions heavier than the
iron group, either from theory or experiment.

\begin{figure}
\includegraphics[width=3.5in]{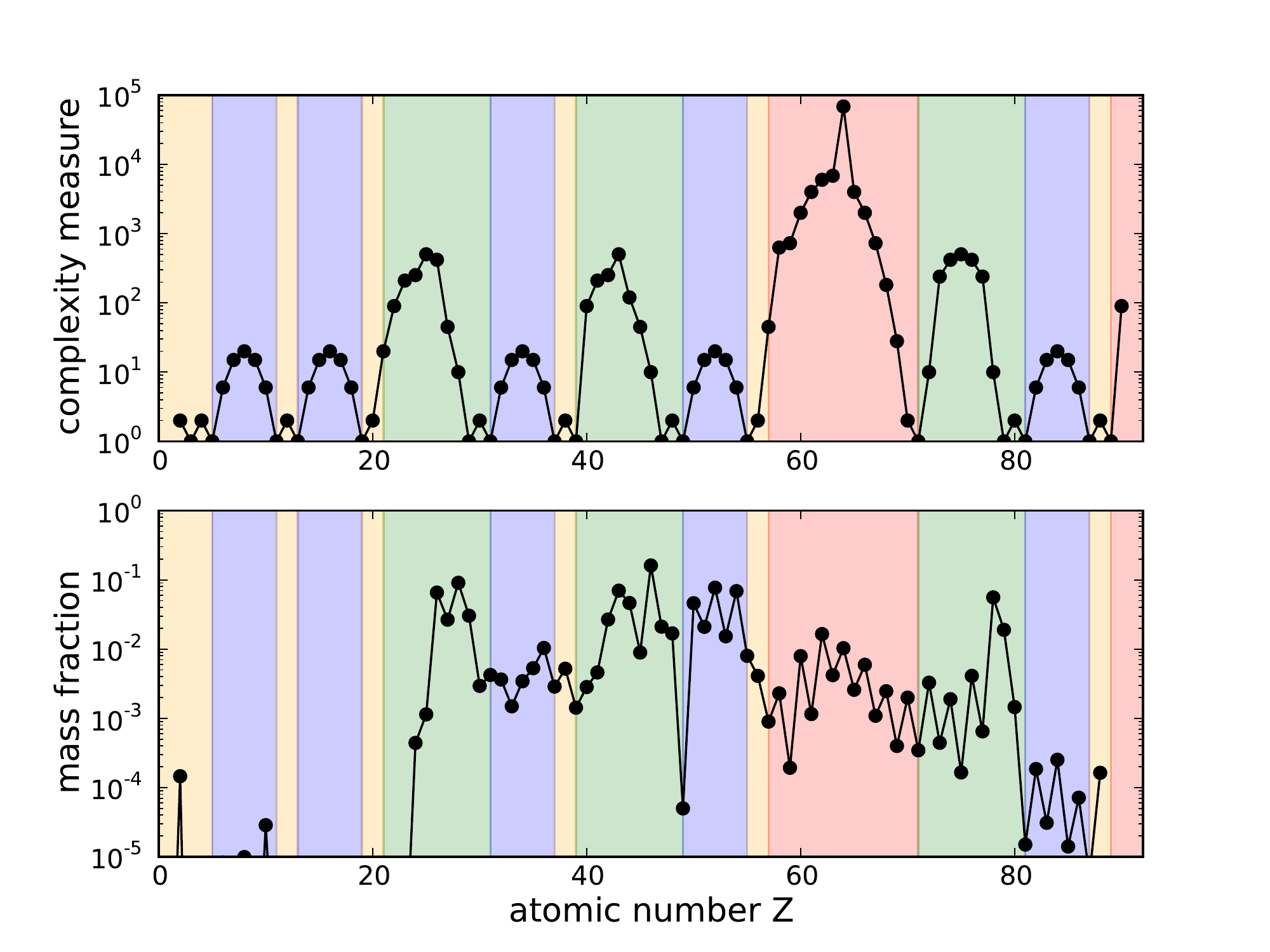}
\caption{ Complexity of the elements (top panel) and their mass
  fractions in the r-process ejecta of neutron star mergers (bottom
  panel).  The top panel plots the number of states in the ground
  configuration for singly ionized ions, as estimated using the simple
  permutation counting of eq.~\ref{eq:complexity}.  The pattern of
  peaks reflects the filling of valance shells, with the color shading
  giving the orbital angular momentum $l$ (yellow = $s$, blue = $p$,
  green = $d$, red = $f$).  The bottom panel plots the mass fractions
  of the tidal tail ejecta from the simulations of \cite{Roberts_2011}.  }
 \label{fig:complexity}
 \end{figure}

Given the ignorance, we might first consider some general expectations
from atomic physics.  The number of strong lines will be larger for
ions with greater complexity -- i.e., with a denser packing of low
lying energy states.  Naively, one might expect higher $Z$ elements to
be more complex than the iron group.  Of course, what matters is not
the total number of electrons, but the number of distinct ways of
distributing valence electrons within the open shells.  A subshell
with orbital angular momentum $l$ has $g = 2(2l +1)$ magnetic
sublevels; one can estimate the number of states in a given electron
configuration by simply counting the permutations of the valence
electrons
\begin{equation}
C = \Pi_i \frac{g_i!}{n_i! (g_i-n_i)!},
\label{eq:complexity}
\end{equation}
where $n_i$ is the number of electrons in the $nl$-orbital labeled
$i$, and the product runs over all open shells in a given
configuration. The different terms and levels (i.e., distinct
combinations of $L,S,J$) derived from these various permutations are
split by electrostatic and fine-structure (e.g. spin-orbit)
interactions.  Equation~\ref{eq:complexity} can thus be used to
estimate the relative number of distinct energy levels of an ion,
while the number of lines (i.e., radiative transitions between levels)
will scale roughly as $C^2$.  Figure~\ref{fig:complexity} plots the
complexity measure $C$ for the ground configurations of singly ionized
ions, where the pattern of $l$ shell filling is clearly seen.

Equation~\ref{eq:complexity} provides immediate insight into the
opacity of r-process ejecta.  Ions with valence shells of higher $l$
are more complex, as are those whose open shells are closer to half
filled.  This is why the iron group, with a nearly half-filled $d$
($l=2$) shell, usually dominates the line opacity in typical
astrophysical mixtures.  Heavy r-process ejecta, however, includes
uncommon species of even greater complexity.  Of particular importance
are the lanthanides ($58 < Z < 70$) and the actinides ($90 < Z < 100$)
which, due to the presence of an open $f$ ($l=3)$ shell, have
complexity measures roughly an order of magnitude greater than the
iron group.  While the actinide series is generally of very low
abundance, the lanthanides may represent several percent of r-process
material by mass.  We will find that these species dominate the total
opacity of NSM ejecta, resulting in opacities orders of magnitudes
greater than previously assumed.

To calculate the ejecta opacity in detail, we need a comprehensive
list of atomic lines.  As almost no data is available for heavy ions,
we turn here to {\it ab initio} atomic structure modeling using the
\AS\ code \citep{Badnell_2011}.  These models determine the
approximate ion energy level structure and the wavelengths and
oscillator strengths of all permitted radiative dipole transitions
(\S\ref{sec:AS}).  Without fine tuning the structure model, the
computed energies and line wavelengths are not exact.  Fortunately,
the effective opacity in an expanding medium is a wavelength average
over many lines.  Because our models reasonably capture the
statistical distribution of levels and lines, they can be used to
derive fairly robust estimates of the pseudo-continuum opacity
(\S\ref{sec:iron}-\S\ref{sec:high_Z}).

Modeling the radiative properties of all high $Z$ ions is a long term
endeavor; we present here initial structure calculations for a few
representative ions selected from the iron group (Fe, Co, Ni), the
lanthanides (Ce, Nd), and a few other heavy d-shell and p-shell ions
(Os, Sn).  The \AS\ line data is then used to calculate the opacity of
expanding ejecta under the assumption of local thermodynamic
equilibrium (LTE).  We show that ions of similar complexity have
similar properties, which allows us to estimate the total opacity of
an r-process mixture based on the representative species
(\S\ref{sec:rp_mix}).

The derived opacities can be input into a multi-wavelength,
time-dependent radiative transfer code to predict the observable
properties of r-process supernovae (\S\ref{sec:spectra}).  We discuss
here the general spectroscopic properties of these transients, while a
companion study explores the broadband light curves and their
dependence on the ejecta properties \citep{Barnes_2013}.  In general,
the high r-process opacities result in light curves that are
significantly broader, dimmer, and redder than  previously
believed.  These results have important implications for observational
strategies to find and interpret the radioactively powered
electromagnetic counterparts to NSMs.

\section{Opacity of Rapidly Expanding Ejecta}

We set the stage by reviewing the physical properties of the material
expected to be ejected in NSMs.  We then describe the nature of the
opacity in such gas, in particular that arising from line interactions
in a rapidly expanding medium.

\subsection{Physical Conditions of the Ejecta}

There are at least two distinct mechanisms by which material may be
ejected in NSMs: 1) During the merger, surface layers may be tidally
stripped and dynamically flung out in ``tidal tails''.  2) Following
the merger, material which has accumulated in a centrifugally
supported disk may be blown off in a neutrino or nuclear driven wind
\citep{Levinson_2006,Surman_2006, Metzger_2008,Metzger_2009}.  The
amount of mass ejected in the tidal tails appears to depends upon many
factors: the NS mass ratio, the equation of state of nuclear matter,
and the treatment of gravity, but simulations give values in the range
$M_{\rm ej} = 10^{-4} - 10^{-1}~\msun$.  A similar amount of mass may
potentially be ejected in the disk wind.  In both cases, the
characteristic velocities are $v_{\rm ej} \approx 0.1-0.3c$, of order
the escape velocity from a NS.

The composition of the material ejected by the two mechanisms is
likely different.  The tidal tail ejecta is initially cold and very
neutron rich (electron fraction $Y_e \sim 0.1$), and should rapidly
assemble into heavy elements ($Z > 50$) through the r-process.  The
conditions in the disk wind are quite different; weak interactions
will drive the material to be less neutron rich ($Y_e \approx 0.5$)
and the entropy will be higher.  This environment is more similar to
the neutrino driven wind from proto-neutron stars in core collapse
supernovae.  It is unclear whether a robust r-process occurs in such a
wind, or whether the distribution only extends to atomic numbers $Z
\sim 50$.  If neutrinos drive $Y_e$ close to 0.5, the composition may
be dominated by radioactive \Nifs\ \citep{Surman_2008}.

Soon after the mass ejection ($\sim 100$'s of seconds) hydrodynamical
and nucleosynthetic processes abate, and the ejecta reaches a phase of
free-expansion.  In the absence of any forces, the velocity structure
becomes homologous -- i.e., the velocity of any mass element is
proportional to radius, $v = r\texp$, where \texp\ is the time since
homology sent in.  The density of the ejecta at 1~day can be estimated
\begin{equation}
\rho_0 \approx \frac{M_{\rm ej}}{(4 \pi/3) v_{\rm ej}^3 \texp^3} \approx 
2.8 \times 10^{-13} \frac{M_{-2}}{v_{0.1}^3 t_d^{3}}~\gcc,
\end{equation}
where $M_{-2} = M_{\rm ej}/10^{-2}~\msun$, $v_{0.1} = v_{\rm
  ej}/0.1c$, and $t_d = \texp/{\rm ~day}$.  In just a day, the density
of the ejecta has dropped by $\sim 20$ orders magnitude from its
original value in the neutron star.

The ejecta material initially cools very effectively by expansion, but
will be reheated by the decay of r-process nuclei.  Radioactive energy
is released in the form of gamma-rays, beta particles, and fission
fragments, which will be thermalized, to various degrees, by
scattering within the ejecta \citep{Metzger_2010}.  The heated
material will radiate, and thermal photons will escape the medium on
the effective diffusion timescale for a homologously expanding medium
\citep{Arnett_1980}
\begin{equation}
t_{\rm d} \sim \biggl[ \frac{ M_{\rm ej} \kappa}{v_{\rm ej} c} \biggr]^{1/2} \sim
 1.7  M_{-2}^{1/2} v_{0.1}^{-1/2} \kappa_{0.1}^{1/2} ~{\rm days},
\end{equation}
where the opacity $\kappa$ of the ejecta has been normalized to a
value $\kappa_{0.1} = \kappa/0.1~{\rm \csg}$, a value appropriate for
iron group elements (but not, we will find, for r-process elements).
This timescale for diffusion sets the duration of the radioactively
powered light curve.

The luminosity near the peak of the light curve will be of order the
instantaneous rate of energy deposition $L \approx M_{\rm ej}
\dot{\epsilon}$, where $\dot{\epsilon}$ is the radioactive energy
released per unit time per unit gram \citep{Arnett_1982}.  Stefan's
law, $L = 4 \pi r^2 \sigma_{\rm sb} T^4$, provides an estimate of the
surface temperature
\begin{equation}
T \approx \biggl[ \frac{M_{\rm ej} \dot{\epsilon}}{\sigma v_{\rm ej}^2 \texp^2} \biggr]^{1/4} \approx
10^4 M_{-2}^{1/4} (v_{0.1} t_d)^{-1/2}~K.
\end{equation}
For times $\texp \gtrsim 1$~day, the characteristic temperatures and
densities of NSM ejecta are thus roughly in the range, $T \sim 10^3 -
10^4$~K and $\rho \sim 10^{-16} - 10^{-12}~{\rm g~cm^{-3}}$.  Under
these conditions, and assuming local thermodynamic equilibrium (LTE),
the gas will be in a low ionization state, being mostly singly or
doubly ionized near the photosphere.

\subsection{Local Thermodynamic Equilibrium}
\label{sec:LTE}

We adopt LTE to compute level populations in this paper, a necessary
approximation given the complexity of the ions involved.  The low
density of NSM ejecta at $\texp \ga 1~$day is not sufficient for
collisional processes alone to establish LTE.  However, in the
optically thick regions below the photosphere, the radiation field
will tend toward a blackbody distribution and radiative transitions
will drive the level populations to their LTE values.  Because the
effective diffusion time and spectral energy distribution (SED) are
mostly set by processes near and below the photosphere, LTE
calculations likely provide a reasonable first approximation.  At late
times $(t \ga 20$~days), when the entire remnant becomes transparent,
LTE will break down at all radii and result in poor SED predictions.

Considering the heavily radioactive environment of NSM ejecta, one may
worry that, even at early times, departures from LTE may be driven by
non-thermal ionization/excitation processes (namely, impacts by fast
electrons that have been Compton scattered by radioactive gamma-rays).
As a rough estimate of the potential effects, we compare the rates for
a bound-bound transition of energy $\Delta E$.  The non-thermal
excitation rate is $R_{\rm nt} \approx f \dot{\epsilon}/ \Delta E$,
where $\dot{\epsilon}$ is the radioactive power released per particle,
and $f$ is the fraction of that power that goes into excitation (as
opposed to ionization or thermalization).  The radiative excitation
rate, assuming a blackbody field, is $R_{\rm bb} = B_{12} B_\nu(T)$
where $B_{12}$ is the Einstein absorption coefficient and $B_\nu(T)$
the Planck function. Using the Einstein relations, the ratio of rates
is
\begin{equation}
\frac{R_{\rm nt}}{R_{\rm bb}} \approx \biggl[ \frac{f \dot{\epsilon} }{ \Delta E  A_{21}} \biggr] 
(e^{\Delta E/kT} - 1),
\label{eq:nt}
\end{equation}
where the Einstein spontaneous emission coefficient $A_{21} \sim 10^8
- 10^9$ for permitted optical transitions.  For r-process ejecta at
1~day, \cite{Metzger_2010} find $\dot{\epsilon} \approx 1~{\rm
  eV~s^{-1}}$ Considering values $\Delta E \sim$~1 eV, $T \approx
5000$~K and $f \sim 1/3$, the ratio $R_{\rm nt}/R_{\rm bb} \approx
10^{-8}$.  Non-thermal transitions are therefore negligible except for
transitions well above the thermal energy ($\Delta E/kT \gtrsim 20)$.
A similar argument can be made for non-thermal ionization.  We
conclude that the radioactive energy deposition will likely not
seriously undermine our LTE assumption, except at late times when the
ejecta becomes rather cold and transparent.

\subsection{Line Expansion Opacity}

The opacity of bound-bound transitions is significantly enhanced by
the high expansion velocities found in supernova and NSM ejecta
\citep{Karp_1977}.  We discuss here the ``expansion opacity''
formalism which wavelength averages the contribution of multiple lines
and treats the line radiative transfer in the Sobolev approximation.

As a photon propagates through the differentially expanding medium,
its wavelength is constantly Doppler shifted with respect to the
comoving frame.  For a homologous (Hubble-like) expansion, this
Doppler shift is always to the red, and proportional to the distance
traveled.  The photon will interact with a line when its comoving
frame wavelength is redshifted into resonance with the line rest
wavelength.  The spatial extant of this interaction, or resonance,
region is set by the intrinsic width of the line profile.  If, for
example, the line width is due to the thermal velocity, $\Delta v_t$,
of ions, the physical size of the resonance region is $\Delta s \sim
\Delta v_t \texp$.

Because the thermal velocities in NSM ejecta are very small ($v_t \sim
1~\kms$) compared to the ejecta velocities ($v_{\rm ej} \approx
10^5~\kms$) the resonance region of a line is tiny compared to the
ejecta scale height, and the matter properties can be taken to be
constant over the region.  This is the essence of the Sobolev
approximation \citep{Sobolev_1960}.  In this limit, the line
extinction coefficient can be analytically integrated to give the
Sobolev optical depth across the resonance region
\begin{equation}
\tau_s = \frac{\pi e^2}{m_e c} \fosc n_{\rm l} \texp \lambda_0,
\end{equation}
where \fosc\ is the oscillator strength and $\lambda_0$ the rest
wavelength of the line.  The Sobolev optical depth is a local quantity
which depends on $n_{\rm l}$, the number density in the lower level of the
transition at the location of resonance.  The probability that a
photon interacts (i.e., is scattered or absorbed) a least once in
traversing the resonance region is simply $1 - e^{-\tau_s}$.

A photon traveling through the expanding medium comes into resonance
with lines one-by-one, sweeping from blue to red.  The effective mean
free path depends not on the strength of any one line, but rather on
the wavelength spacing of strong ($\tau_s \gg 1$) lines, which can be
quantified as follows.  Say that within some wavelength region
$(\lambda_i, \lambda_i + \dls)$ we have $N$ strong lines.  The spacing
between the lines is, on average, $\dls/N$ and the velocity gradient
which Doppler shifts a photon from one line to the next is $\dvs/c =
\dls/ \lambda_i N$.  If homologous expansion holds, the distance a
photon travels between line interactions is then $l_{\rm mfp} = \dvs
\texp$.  This is an estimate of the mean free path, while the inverse
quantity, $l_{\rm mfp}^{-1}$, defines the matter extinction
coefficient (units cm$^{-1}$)
\begin{equation}
\aex \approx \frac{1}{l_{\rm mfp}}
\approx  \frac{N }{\dls}  \frac{\lambda_i}{ c \texp},
\label{eq:simple_ex}
\end{equation}
where the bin size \dls\ can be chosen arbitrarily to average over a
reasonable number of lines.

A formal derivation along these lines was introduced by
\cite{Karp_1977} to estimate the extinction coefficient in an
expanding medium.  We use here the slightly modified expression
developed by \cite{Eastman_1993}
\begin{equation}
\aex(\lambda) = \frac{1}{c \texp}  \sum_i \frac{\lambda_i}{\dls } 
 [1 - e^{-\tau_i} ],
 \label{eq:alpha_ex}
\end{equation}
where the sum runs over all lines in the wavelength bin \dls.
Eq.~\ref{eq:alpha_ex} takes into account the cumulative effect of many
weak lines; in the case that all lines are optically thick ($\tau_s
\gg 1$), it reduces to the simple estimate eq.~\ref{eq:simple_ex}.

An interesting property of the expansion opacity is the dependence on
ion density, which appears only in the Sobolev optical depth,
$\tau_s$.  In the limit that all lines are weak, the density
dependence is linear. However, in the opposite limit where all lines
are very optically thick, the extinction coefficient is independent of
density.  In our actual calculations, the density dependence (which is
set by the statistical distribution of line optical depths) is weak,
roughly logarithmic.  As a consequence, certain ions can make a
significant contribution to the opacity even when their abundance is
very low.

The expansion opacity (units \csg) of bound-bound transitions is
related to $\alpha_{\rm ex}$ by
\begin{equation}
\kappa_{\rm ex}(\lambda) = \frac{\alpha_{\rm ex}}{\rho} = \frac{1}{c \texp \rho}  \sum_i \frac{\lambda_i}{\Delta \lambda } 
 [1 - e^{-\tau_i} ]
 \label{eq:ex_opacity}
\end{equation}
Given the weak dependence of $\alpha_{\rm ex}$ on density, the line
expansion opacity actually increases as the density drops.

\subsection{Applicability of the Sobolev Approximation}
\label{sec:sobolev}

At least three conditions must be met for the Sobolev approximation,
which underlies the expansion opacity expression
eq.~\ref{eq:ex_opacity}, to be valid.  The first, already mentioned,
is that the thermal velocity of the ions (presumed to set the
intrinsic line widths) must be significantly smaller than the velocity
scales over which the ejecta properties vary.  For NSM ejecta, the
ratio is $v_t/v_{\rm ej} \approx 10^{-5} \ll 1$, which assures the
applicability of the narrow line limit.

A second condition, relevant for time-varying environments, is that
the time photons spend within a resonance region be short compared to
the timescale over which the ejecta properties vary (e.g., the
expansion timescale).  For strong lines, a photon may scatter multiple
times within the resonance region before finally being redistributed
to the edge of the line profile and escaping (i.e., redshifting past)
the transition.  In the Sobolev formalism, the probability that a
photon escapes the line after a scatter is
\begin{equation}
\beta = \frac{1 - e^{-\tau_s}}{\tau_s} \approx  \tau_s^{-1}~{\rm for~\tau \gg 1},
\end{equation}
and the average number of scatters in a thick line is $N_{\rm sc} \sim
1/\beta$.  Assuming the distance traveled between scatters is, on
average, $v_{\rm t}\texp$, the time spent in the resonance region is
$t_{\rm sc} \sim N_{\rm sc} v_{\rm t} \texp/c$.  The condition $t_{\rm
  sc} < \texp$ sets a limit on the optical depth of the line $\tau_s
\la c/v_{\rm t} = 3\times 10^5 v_{t,1}$, where $v_{t,1} = v_t/1~\kms$.
In practice, optical depths $> 3\times 10^5$ are regularly
encountered, in particular for resonance lines.  In most cases,
however, fluorescence provides an avenue for escape.  The probability
of de-excitation to a lower level is suppressed by a factor of
$\beta$, and so it is likely that the ion will eventually de-excite
via a cascade through multiple low $\tau_s$ transitions.  This
generally evades the problem of extended line trapping, except perhaps
for those few transitions in which fluorescence is not possible.

A third condition is that the intrinsic profiles of strong lines must
not, in general, overlap, as this would introduce a coupling of the
radiative transport between lines.  In particular, a photon that
escapes from the red edge of one line will have an enhanced
probability of escaping a second overlapping line.  This invalidates
the sum in eq.~\ref{eq:ex_opacity} which assumes an independent
interaction probability for each line.  Overlap of weak lines (which
are extremely numerous in our calculations) is common, however this
likely does not introduce any serious error, as the $\tau_s$
dependence is linear when $\tau_s \ll 1$.  Moreover, the opacity is
usually dominated by the strong lines.  Occasional overlap of strong
lines is inevitable, and may moderately reduce the expansion opacity
at certain wavelengths.  The entire Sobolev formalism, however,
becomes inapplicable when the wavelength spacing of strong lines,
$\Delta\lambda/N$, becomes comparable to the intrinsic (e.g., thermal)
width $\Delta \lambda_{\rm t} = \lambda_0 (v_{\rm t}/c)$ of the lines.
From eq.~\ref{eq:ex_opacity} we can define an opacity when such
``saturation" occurs
\begin{equation}
\kappa_{\rm sat} =  \frac{\lambda_0}{\Delta \lambda_{\rm t} }\frac{1}{\rho c \texp} 
= \frac{1}{\rho v_{\rm t} t_{10}} \approx 10^3
~\rho_{-13} t_{1}^{-1} v_{\rm t,1}^{-1}~\csg.
\label{eq:kappa_sat}
\end{equation}
When $\kappa_{\rm ex} > \kappa_{\rm sat}$, strong line overlap is the
norm and the Sobolev expansion opacity formalism can no longer be
trusted to return reasonable values.  Under some conditions, and at
certain wavelengths, we will find that our calculated r-process
opacities approach or exceed saturation, such that this issue may be a
serious concern.

\subsection{Other sources of Opacity}

Other potential sources of opacity include free-free (i.e.,
bremsstrahlung), bound-free (i.e., photoionization), and electron
scattering.  In NSM ejecta, none of these turn out to be important
compared to bound-bound.  For example, the wavelength independent
electron scattering opacity is given by
\begin{equation}
\kappa_{\rm es} = \frac{\bar{x} \sigma_t }{\bar{A} m_p} \approx 0.4 \biggl( \frac{\bar{x}}{\bar{A}}\biggr)  \csg,
\end{equation}
where $\bar{x}$ is the mean ionization fraction and $\bar{A}$ the mean
atomic weight of the ions.  For NSM ejecta comprised of lowly ionized
($\bar{x} \sim 1$) heavy elements $(\bar{A} \sim 130$), $\kappa_{\rm
  es}$ is a factor $\bar{x}/\bar{A} \sim 10^{-2}$ smaller than the
typical value for ionized hydrogen, and much less than the r-process
line opacity at all wavelengths of interest.

The free-free opacity for a gas in ionization state $\bar{x} \sim 1$
is given approximately by \citep[e.g.,][]{rybicki_lightman}
\begin{equation}
\kappa_{\rm ff} = 0.005 \frac{\bar{x}^3}{A^2} \rho_{-13} T_{4}^{-1/2} (\lambda/1 {\rm \mu m})^{3}~\csg,
\end{equation}
where we have set the correction for stimulated emission and the Gaunt
factor to unity.  For the low densities and $\bar{x}/\bar{A}$ values
found in NSM ejecta, the free-free opacity at the relevant wavelengths
is negligible ($\kappa_{\rm ff} \approx 10^{-6}$~\csg).

Finally, the opacity due to a bound-free transition for photons at the
threshold energy (where the cross-section is largest) for some excited
level of an ion is given by
\begin{equation}
\kappa_{\rm bf} = \frac{\sigma_0}{\bar{A} m_p} \frac{e^{-\Delta E/kT}}{Z(T)}\,,
\end{equation}
where $\Delta E$ is the excitation energy of the level and $Z(T)$ the
partition function (LTE is assumed). In order for the bound-free
transition to apply to optical/infrared photons, the level must be a
highly excited state, $\Delta E \gtrsim \chi - 2$~eV, where $\chi$ is
the ionization potential.  As a representative estimate of the opacity
at $T = 5000$~K, we consider a singly ionized heavy element with $A =
120, \Delta E \sim 8$~eV and $Z(T) = 20$ and adopt the hydrogenic
value $\sigma_0 \approx 6 \times 10^{-18}~{\rm cm^2}$.  The resulting
opacity at threshold is also very small, $\kappa_{\rm bf} \approx 1.3
\times 10^{-5}~\csg$, due mainly to the Boltzmann factor. At
ultraviolet wavelengths ($\lambda \lesssim 1000$~\AA) the bound-free
opacity may actually dominate, since for low lying levels ($\Delta E
\approx 0$) one finds $\kappa_{\rm bf} \approx 10^3~\csg$.

\section{Atomic Structure Calculations}
\label{sec:AS}

\begin{figure}
\includegraphics[width=3.5in]{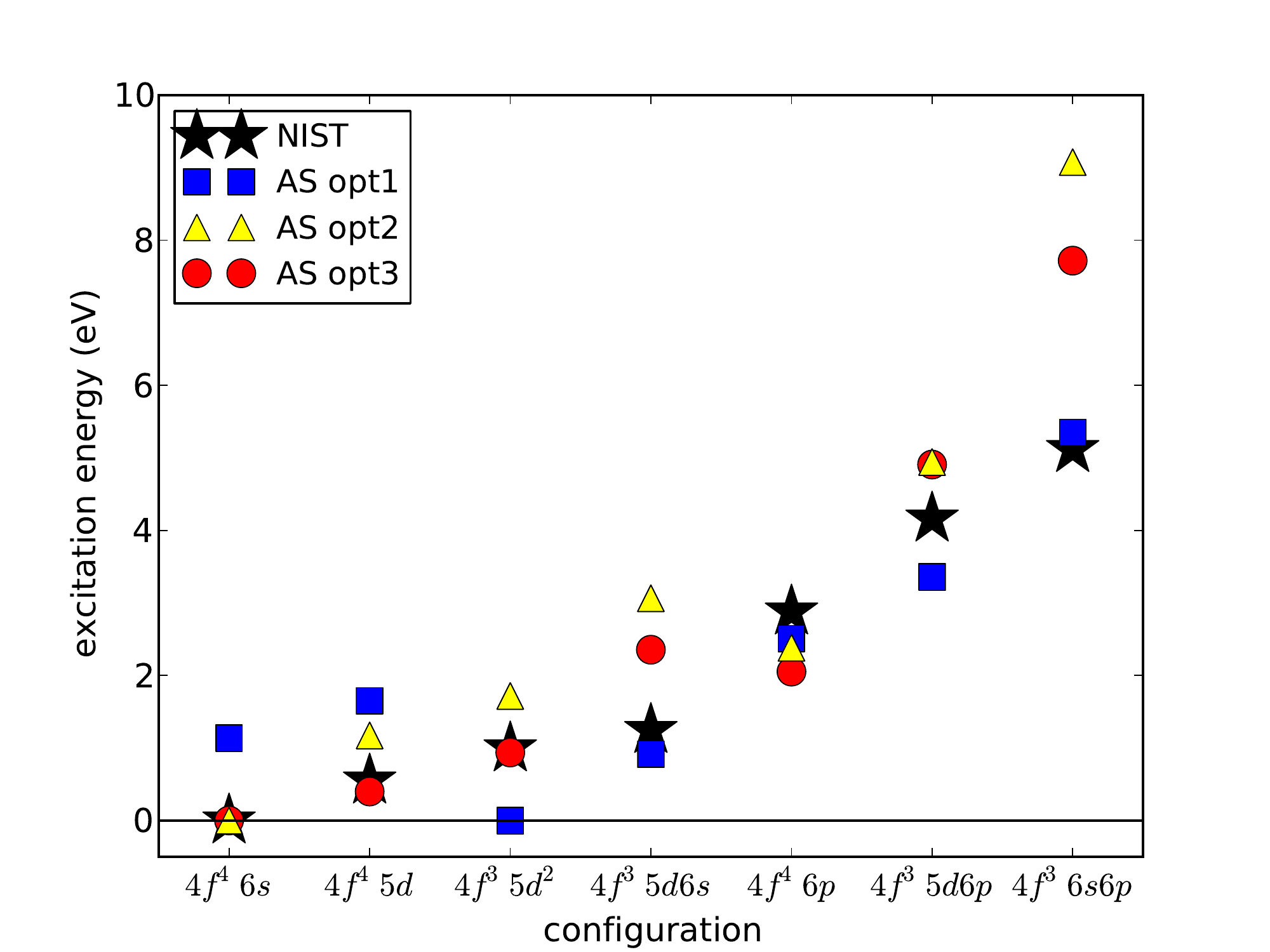}
\caption{Atomic structure model calculations of the excitation energy
  of the lowest level of NdII electron configurations.  Circles denote
  the results from \AS\ obtained under various optimization approaches
  (described in text).  Stars denote the experimental energies from
  NIST.
 \label{fig:Nd_energies}}
 \end{figure}


\begin{table*}
\caption{Autostructure atomic structure models \label{tab:configs}}
\begin{small}
\begin{center}
\begin{tabular}{lcccc}
        \tableline
	\tableline
Ion & Configurations included & \# levels & \# lines & $\chi$ (eV)\\
        \tableline
FeI & ${\bf 3d^6 4s^2}, 3d^7 4s, 3d^6 4s4p, 3d^7 4p, 3d^7 4d, 3d^7 4f, 3d^7 5s, 3d^7 5p, 3d^7 5d, 3d^6 4s 4d$ & 1784  & 326,519 & 7.90 \\
FeII &  ${\bf 3d^6 4s}, 3d^7, 3d^6 4p, 3d^6 4d, 3d^6 4f, 3d^6 5s, 3d^6 5p, 3d^6 5d, 3d^5 4s^2, 3d^5 4s 4p$ & 1857  & 355,367 &16.18 \\
FeIII &  ${\bf 3d^6}, 3d^5 4s, 3d^5 4p, 3d^5 4d, 3d^5 4f, 3d^5 5s, 3d^5 5p, 3d^5 5d, 3d^4 4s 4p$ & 2050 & 420,821 & 30.65 \\
FeIV & ${\bf 3d^5}, 3d^4 4s, 3d^4 4p, 3d^4 4d, 3d^4 4f, 3d^4 5s, 3d^4 5p, 3d^4 5d$ & 1421 & 217,986 &54.91 \\
CoI  & ${\bf 3d^7 4s^2}, 3d^8 4s, 3d^7 4s 4p, 3d^9, 3d^8 4p, 3d^8 4d, 3d^8 5s, 3d^7 4s 4d, 3d^7 4s 5s$ & 778 & 62,587 & 7.88 \\
CoII & ${\bf 3d^8}, 3d^7 4s, 3d^6 4s^2, 3d^7 4p, 3d^6 4s 4p, 3d^7 5s, 3d^7 4d$ & 757 & 58,521 & 17.08 \\
CoIII & ${\bf 3d^7}, 3d^6 4s, 3d^6 4p, 3d^6 4d, 3d^6 5s$ & 601 & 34,508 & 33.50\\
CoIV & ${\bf 3d^6}, 3d^5 4s, 3d^5 4p, 3d^5 4d, 3d^5 5s$ & 728 & 48,254 &51.27  \\
NiI & ${\bf 3d^8 4s^2}, 3d^{10}, 3d^8 4s 4p, 3d^9 4s, 3d^9 4p, 3d^9 4d, 3d^9 4f, 3d^9 5s, 3d^9 5p, 3d^9 6s$ & 174 & 2,776 & 7.64\\
NiII & ${\bf 3d^9},  3d^8 4s, 3d^8 4p, 3d^8 4d, 3d^8 4f, 3d^8 5s, 3d^8 5p, 3d^8 6s, 3d^7 4s4p, 3d^7 4s^2$ & 520 & 25,496 & 16.18 \\
NiIII & ${\bf 3d^8},  3d^7 4s, 3d^7 4p, 3d^7 4d, 3d^7 4f, 3d^7 5s, 3d^7 5p, 3d^7 6s, 3d^6 4s^2$ & 1644 & 61,108 & 35.19 \\
NiIV & ${\bf 3d^7},  3d^6 4s, 3d^6 4p, 3d^6 4d, 3d^6 4f, 3d^6 5s, 3d^6 5p, 3d^6 6s, 3d^5 4s4p, 3d^5 4s^2$ & 751 & 258,305 & 54.92 \\
NdI & ${\bf 4f^4 6s^2}, 4f^3 5d 6s^2, 4f^4 5d 6s, 4f^4 5d^2, 4f^3 5d 6s 6p, 4f^4 5d 6p$  & 18104 & 24,632,513 & 5.52\\
NdII & ${\bf 4f^4 6s}, 4f^4 5d, 4f^4 6p, 4f^3 5d^2, 4f^3 5d 6s, 4f^3 5d 6p, 4f^3 6s 6p $ & 6888 & 3,873,372 & 10.7\\
NdIII & ${\bf 4f^4}, 4f^3 5d, 4f^3 6s, 4f^3 6p, 4f^2 5d^2, 4f^2 5d 6s, 4f 5d^2 6s$ & 1650 &232,715 & 22.14\\
NdIV & ${\bf 4f^3}, 4f^2 5d, 4f^2 6s, 4f^2 6p$ & 241 & 5780 & 40.4\\
CeII & ${\bf 4f 5d^2}, 4f 5d 6s, 4f^2 6s, 4f^2 5d, 4f 6s^2, 4f 5d 6p, 4f^2 6p, 5d^3, 4f 6s 6p,4f^3$ & 5,637 & 
4,349,351 & 10.8\\
CeIII & ${\bf 4f 5d}, 4f 6s, 5d^2, 4f 6p, 5d 6s$ & 3,069  & 868,640 & 20.19 \\
OsII & ${\bf 5d^6 6s}, 5d^6 5f, 5d^6 5g, 5d^6 6s, 5d^6 6p, 5d^6 6d, 5d^6 6f, 5d^6 6g $ & 3271 & 1,033,972 & 17.0\\
SnII & ${\bf 5s^2 5p}, 5s^2 4f,  5s^2 5d, 5s^2 6s, 5s^2 6p, 5s 5p^2, 5s 5p 6s, 5s 5p 6p$ & 47 & 371  &14.63 \\
		\tableline
\end{tabular}
\end{center}
\end{small}
\label{tab:config}
\end{table*}

 
To estimate the radiative data for high-Z elements, we used the
\AS\ program \citep{Badnell_2011}. This code was used previously to
calculate data up to Ni for the updated opacities of the Opacity
Project \citep{Badnell_2005}. Recent developments \citep{Badnell_2012}
have enabled it to be used to make extensive calculations of radiative
(and autoionization) rates for a half-open f-shell.  \AS\ calculates
the approximate level energy structure of ions, and all relevant
radiative transition rates, given a user specified set of electron
configurations.  The many-electron quantum mechanical problem is
treated using a multi-configuration wavefunction expansion with a
Breit-Pauli Hamiltonian.  We used the kappa-averaged relativistic
wavefunction option as introduced by \cite{Cowan_1976}.  The radial
orbitals were determined using a Thomas-Fermi-Dirac-Amaldi potential.
As the standard $LS$-coupling scheme breaks down for high $Z$
elements, we adopted a level-resolved intermediate coupling scheme.

We used the NIST \citep{NIST} atomic database to identify the electron
configurations corresponding to the ground and low-lying states of
each ion.  For several of the high-Z ions, the NIST data appeared to
be incomplete, and we included additional configurations suspected to
be relevant. Table~\ref{tab:config} lists the configurations used for
each ion.  For the lanthanides, the highest $nl$ orbital we considered
was the $6p$ one.  We experimented with including configurations
generated by electron promotion to higher orbitals (e.g., $n=7,8$),
however these typically produced highly excited levels not
significantly thermally populated under the relevant physical
conditions.  As we did not did not notice any large effects on the
opacities, we omit these configurations from our final calculations,
although more exhaustive explorations of configuration space are
certainly warranted.

\AS\ includes a dimensionless radial scaling parameter for each
$nl$-orbital, which must be optimized to establish a realistic level
structure for low-charge ions.  The optimization consists of varying
the scaling parameters so as to minimize a user specified weighted sum
of eigenenergies.  The closed-shell core cannot be excluded from the
structural optimization of complex heavy near-neutral ions because of
the strength of core polarization effects on the valence orbitals
\citep{Palmeri_2000}.  We therefore used a single common variational
scaling parameter for all closed-shell orbitals, but varied the
parameters of the valence orbitals independently.  This {\it ab
  initio} optimization procedure does not require any observed
energies.  Thus, it is ideally suited to situations such as the
present one where the observed data is at best sparse.

We explored several strategies for optimization.  The first -- which
we label {\it opt1} -- was to simultaneously vary the scaling
parameters for all included (core-plus-valence) orbitals and to
minimize the equally weighted sum of all energy levels included by the
configuration expansion.  This has the advantage of not biasing the
structure towards any given configuration(s), which is valuable given
that we seek radiative data for many excited levels.  The disadvantage
is that it gives no due preference to the ground state, and thus does
not always predict a ground state configuration in agreement with what
is deduced experimentally.  Figure~\ref{fig:Nd_energies} shows the
energy level structure for NdII, where it is seen that the {\it opt1}
optimization predicts the wrong ground configuration.  While for many
applications this would be a fatal flaw, in the present context we
expect several low lying configurations to be significantly populated,
in which case the mean line opacity may be less sensitive to the exact
configuration ordering.

We considered a second strategy ({\it opt2}) whereby the above
optimization was first applied to only those orbitals included in the
ground configuration.  These scaling parameters were then fixed, and a
second optimization was carried out varying the parameters of all
remaining orbitals. This method usually produced the correct ground
state configuration.  The energies of the excited levels were also
close to but a bit higher than the available NIST values, and overall
not as good as those found using the {\it opt1} approach
(Figure~\ref{fig:Nd_energies}).

The model structure can be further refined by iteratively adjusting
the scaling parameters by hand.  We attempted this for NdII, guided by
the trends found in the {\it opt1} and {\it opt2} calculations.  An
improved solution was found ({\it opt3}) which reproduced the ground
and first two excited level energies almost exactly.  Further
iterations could presumably improve the result, but this sort of
manual alignment is time consuming, and more art than science.  We
only attempted this {\it opt3} approach only for NdII, which is the
most important ion for our r-process light curve calculations.

\section{Iron Group Opacities}
\label{sec:iron}

\begin{figure*}[t]
\includegraphics[width=3.5in]{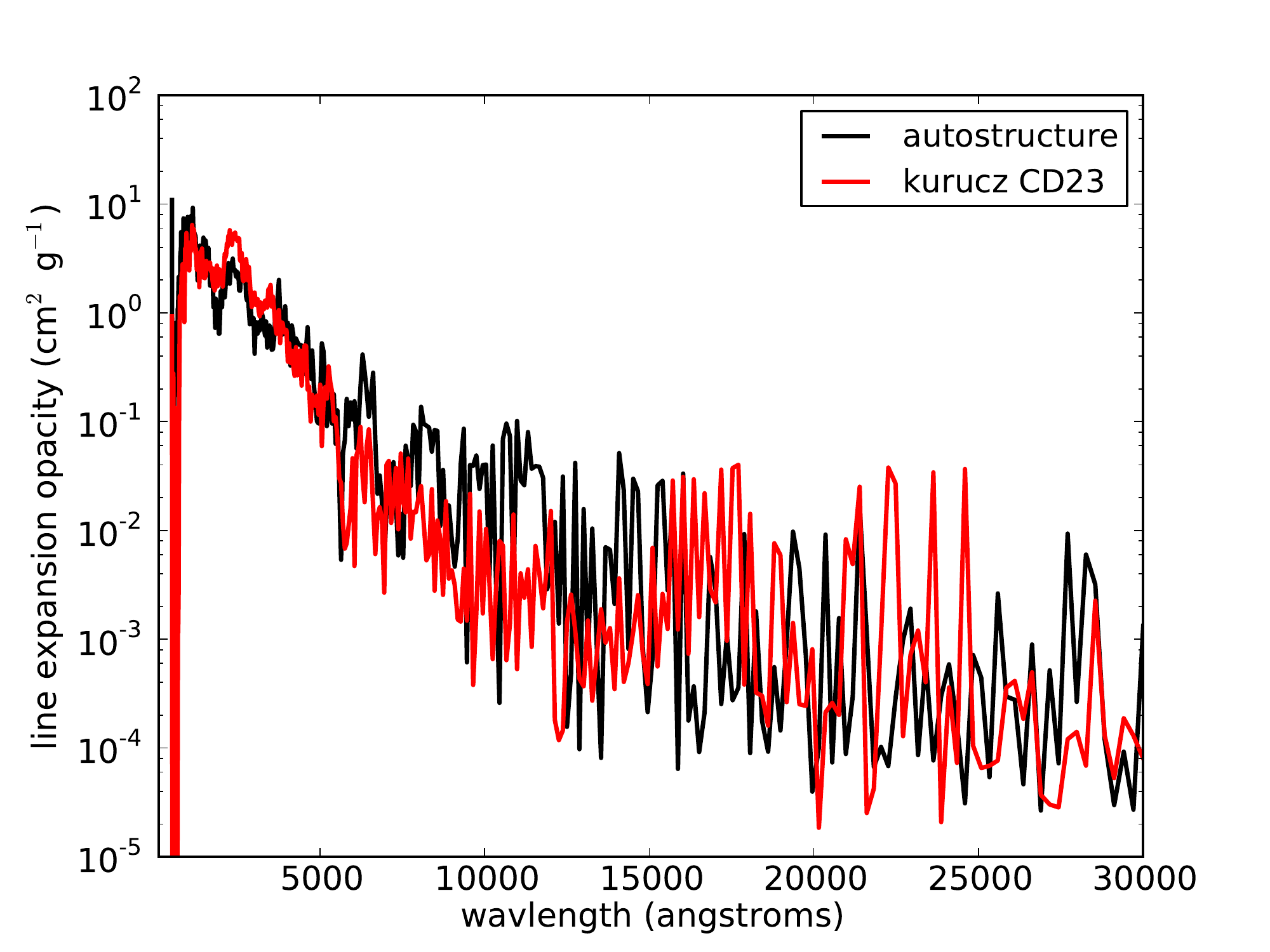}
\includegraphics[width=3.5in]{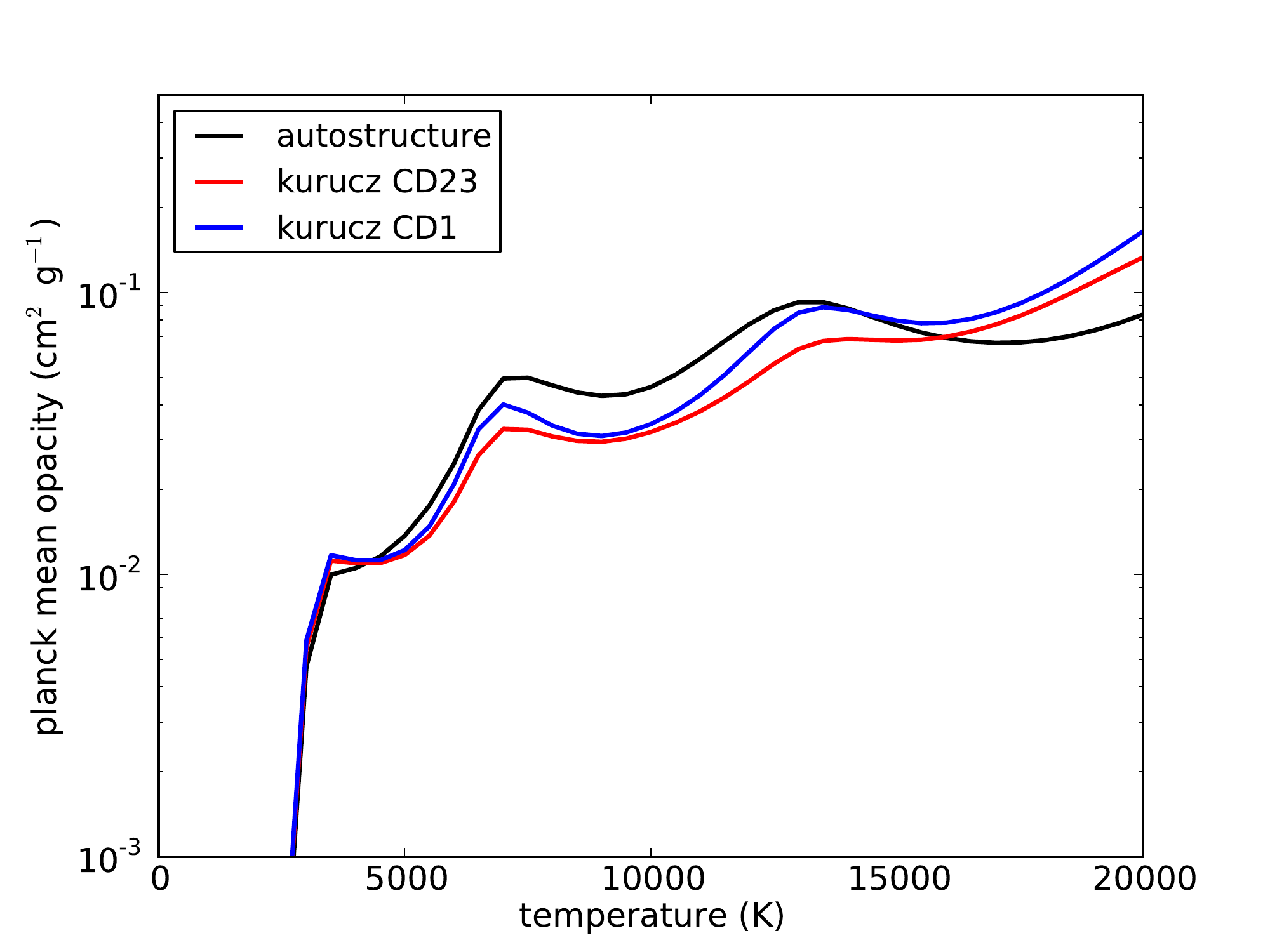}
\caption{Calculated line expansion opacities for a mixture of iron
  group elements (10\% Ni, 80\% Co, 10\% Fe) representative of decayed
  \Nifs.  The right panel plots the Planck mean opacity versus
  temperature, while the left panel plots the wavelength dependence of
  the opacity at a specific temperature, $T = 5000$~K. The opacities
  calculated using our \AS\ derived line data generally agree with
  those using the Kurucz linelists.  The calculations assume $\rho =
  10^{-13}$~\gcc, $\texp = 10$~days, and a wavelength binning $\Delta
  \lambda = 0.01 \lambda$.  }
 \label{fig:feg_mean_opac}
 \end{figure*}

\subsection{Comparison to Kurucz Line Data}
 
The atomic properties of $Z < 30$ ions are reasonably well known based
on experiment and previous structure modeling.  In particular,
R.~Kurucz has generated extensive line lists, including CD23, ($\sim
500,000$, \cite{Kurucz_CD23}), and CD1 ($\sim 42$ million lines,
\cite{Kurucz_CD1}).  These lists (which are dominated by iron group
lines) have been derived from atomic structure calculations using the
Cowan code \citep{Cowan_1981} which have been iteratively tuned to
reproduce the extensive observed experimental level energies
\citep{Kurucz_Bell_1995}.  Supernova modelers have used the Kurucz
data to successfully model the optical light curves and spectra of
observed (iron-rich) SNe~Ia \citep[e.g.,][]{Kasen_2009,Sim_2010} which
suggests that, for the iron group, the Kurucz line data can be taken
to be reasonably accurate and complete.
 
To validate our {\it ab initio} \AS\ line data against the
observationally constrained data of Kurucz, we ran structure models
for the first 4 ionization stages of Fe, Co, and Ni, using the
electron configurations listed in table~\ref{tab:configs}.  Unlike
Kurucz, we made no attempt (beyond our {\it ab initio} {\it opt1}
optimization scheme) to tune the model, and our calculated level
energies can differ from the experimental values by factors of 2 or
more.  Nevertheless, we find that our derived iron group expansion
opacities are in good agreement with those of Kurucz
(Figure~\ref{fig:feg_mean_opac}). Our Planck mean opacities differ
from those of Kurucz by only $\sim 30\%$ over the temperature range
$1000-20,000$~K, and the wavelength dependence of the opacity is
quantitatively similar, with the opacity rising sharply to the blue.
The good agreement indicates that our \AS\ calculations capture the
statistical properties of the lines, even if the individual energy
levels and line wavelengths may be inaccurate.

\subsection{Application to Supernova Modeling}
\label{sec:sn1a}

 \begin{figure*}
\includegraphics[width=3.5in]{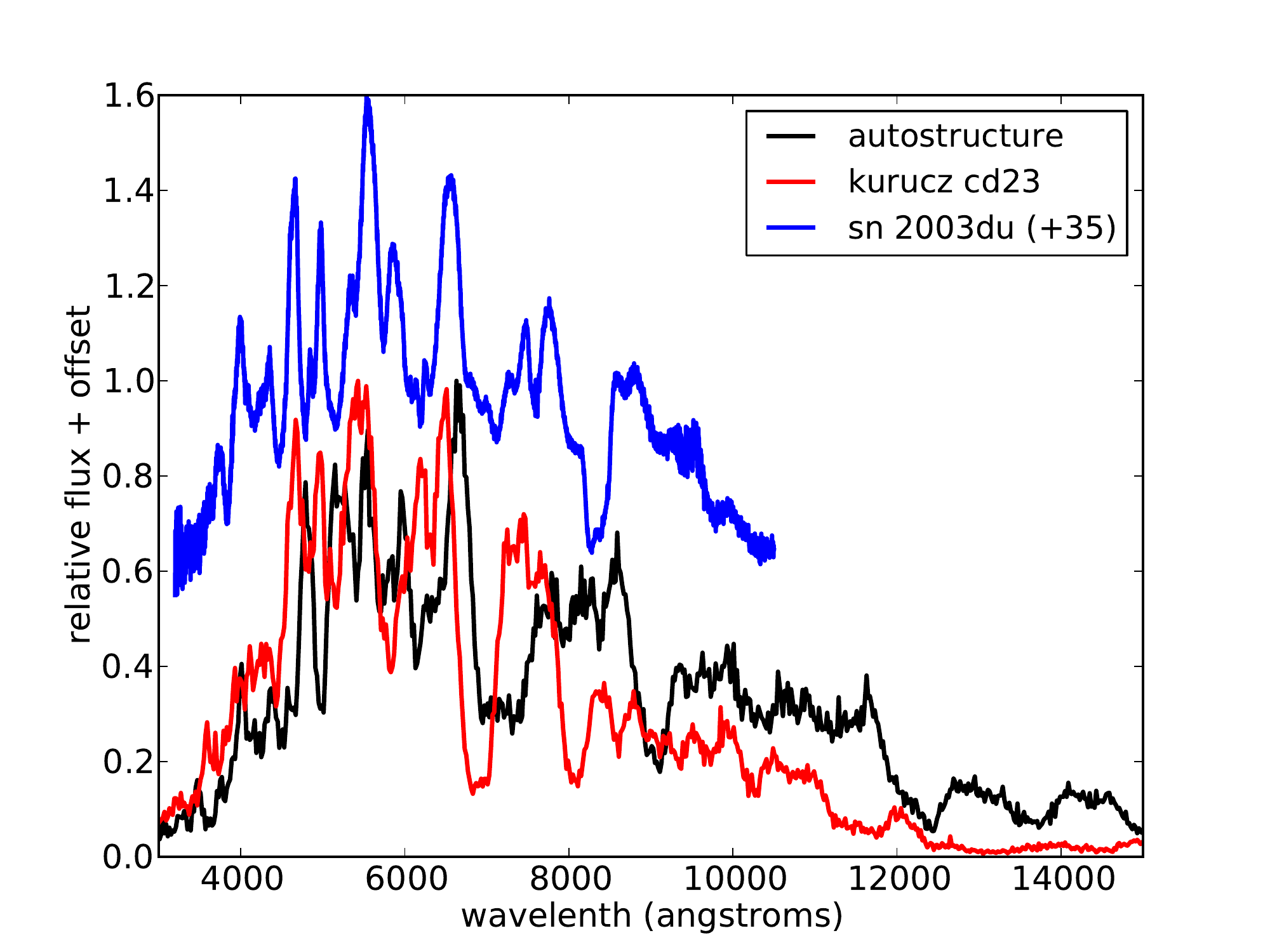}
\includegraphics[width=3.5in]{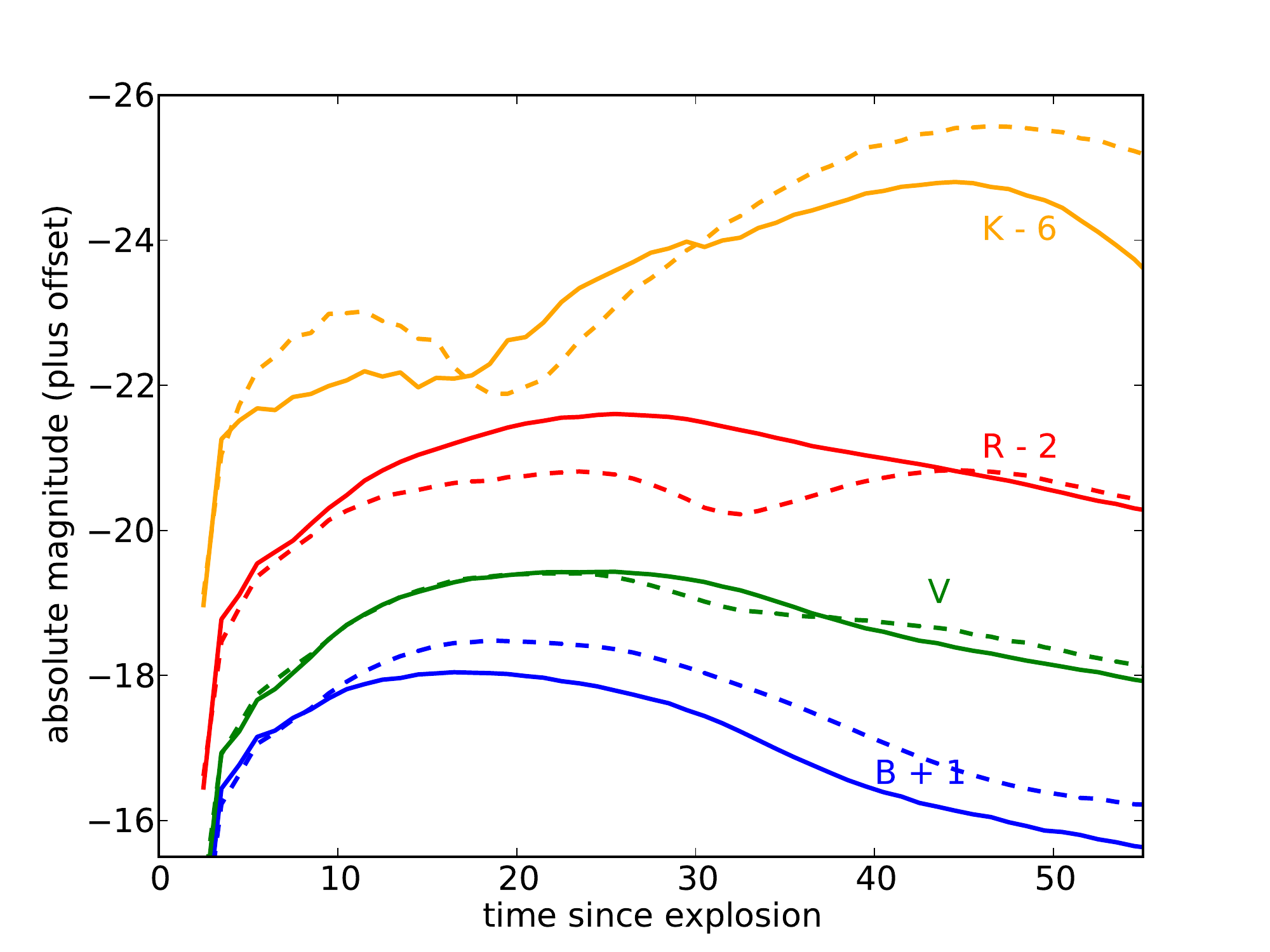}
\caption{Synthetic light curves and spectra of a model consisting of
  pure \Nifs\ ejecta, computed with different line data.  The left
  panel shows that the SED (at 50 days after merger) calculated using
  the \AS\ line data resembles that obtained using the Kurucz
  data. Both resemble the spectrum of the Type~Ia SN~2003du observed
  34 days after peak ($\sim 52$ days after explosion).  The right
  panel compares the broadband light curves of the model calculated
  using the \AS\ line data (solid lines) and the Kurucz CD23 linelist
  (dashed lines).  }
 \label{fig:feg_spec}
 \end{figure*}
 
To demonstrate how our \AS\ derived iron group opacities perform in a
real transport calculation, we calculated synthetic light curves and
spectra of a simple SN~Ia model.  As a numerical ``thought"
experiment, we pretended that our understanding of SNe~Ia was as
rudimentary as it is for NSM outflows, and that our only expectation
was that a carbon/oxygen star was burned to nuclear statistical
equilibrium (NSE).  We thus constructed a spherically symmetric ejecta
model consisting of uniform density, initially pure \Nifs\ with a
total mass of $1~\msun$ and kinetic energy of $10^{51}$~ergs, roughly
the nuclear energy released in burning the C/O to NSE.  Of course,
real SN~Ia are not homogenous, and in addition to \Nifs\ are observed
to contain a significant amount of intermediate mass elements (IMEs,
Si, S, Ca).

We calculated synthetic observables of this ejecta model using the
time dependent, multi-wavelength radiation transport code
\Sedona\ \citep{Kasen_MC}, and assuming that the level populations
were in LTE.  Figure~\ref{fig:feg_spec} shows that, despite the
simplistic nature of the ejecta model, the broadband light curves
qualitatively resemble those of observed SNe~Ia, peaking at a B-band
magnitude around -19 about 20 days after explosion.  Overall, the
light curves calculated using our \AS\ derived linelist are rather
similar to those calculated using the Kurucz linelist, although
differences up to 1 magnitude are seen at some epochs.

Figure~\ref{fig:feg_spec} shows that the model spectrum (at 50 days
after explosion) derived from the \AS\ linelist is also similar to
that using Kurucz.  Both calculations resemble the SED of an observed
SN~Ia.  The \AS\ model does not reproduce the positions of most
spectral features, which is to be expected given that the line
wavelengths are only approximate.  Even the Kurucz calculation fails
to reproduce every observed spectral feature, as the underlying ejecta
model did not include the IMEs present in real SNe~Ia.

These results indicate that line data derived from our \AS\ models can
be used to predict supernova SEDs (but not line features) with some
reliability.  The general agreement of our synthetic observables with
those of real SNe~Ia suggests that -- even with very crude knowledge
of the underlying ejecta structure -- we may still be able to predict
the light curve and colors of radioactive transients to a reasonable
level of accuracy.

\section{High $Z$ Opacities}
\label{sec:high_Z}

\begin{figure}
\includegraphics[width=3.5in]{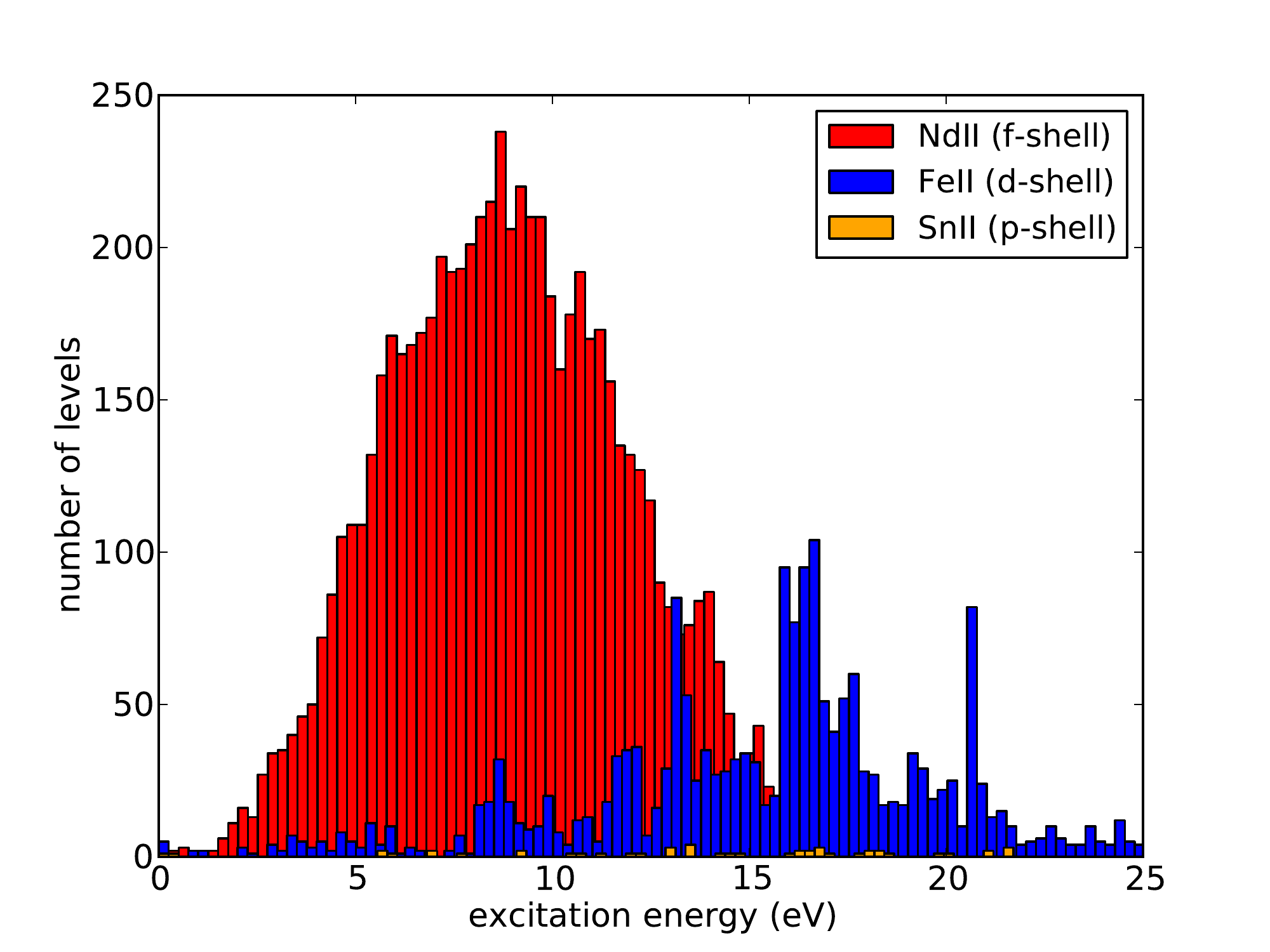}
\caption{ Histogram of the number of atomic levels versus level energy
  (bin size = 0.25~eV) in our \AS\ models, which illustrates the much
  greater complexity of the lanthanide neodymium (with an open
  f-shell) as compared to iron (open d-shell) and tin (open p-shell).
\label{fig:lev_dens}}
 \end{figure}

We have calculated structure models for several elements beyond the
iron group, including tin (Sn, $Z=50$, p-shell), cerium (Ce, $Z=58$,
f-shell), neodymium (Nd, $Z=60$, f-block), and osmium (Os, $Z=76$,
d-shell).  These species were chosen to sample different blocks on the
periodic table corresponding to valence shells of different orbital
angular momentum.  The total number of atomic levels/lines determined
by the structure models are listed in Table~1 and illustrated in
Figure~\ref{fig:lev_dens}, and are generally consistent with the
simple complexity estimates of \S\ref{sec:intro}.

\begin{figure}
\includegraphics[width=3.5in]{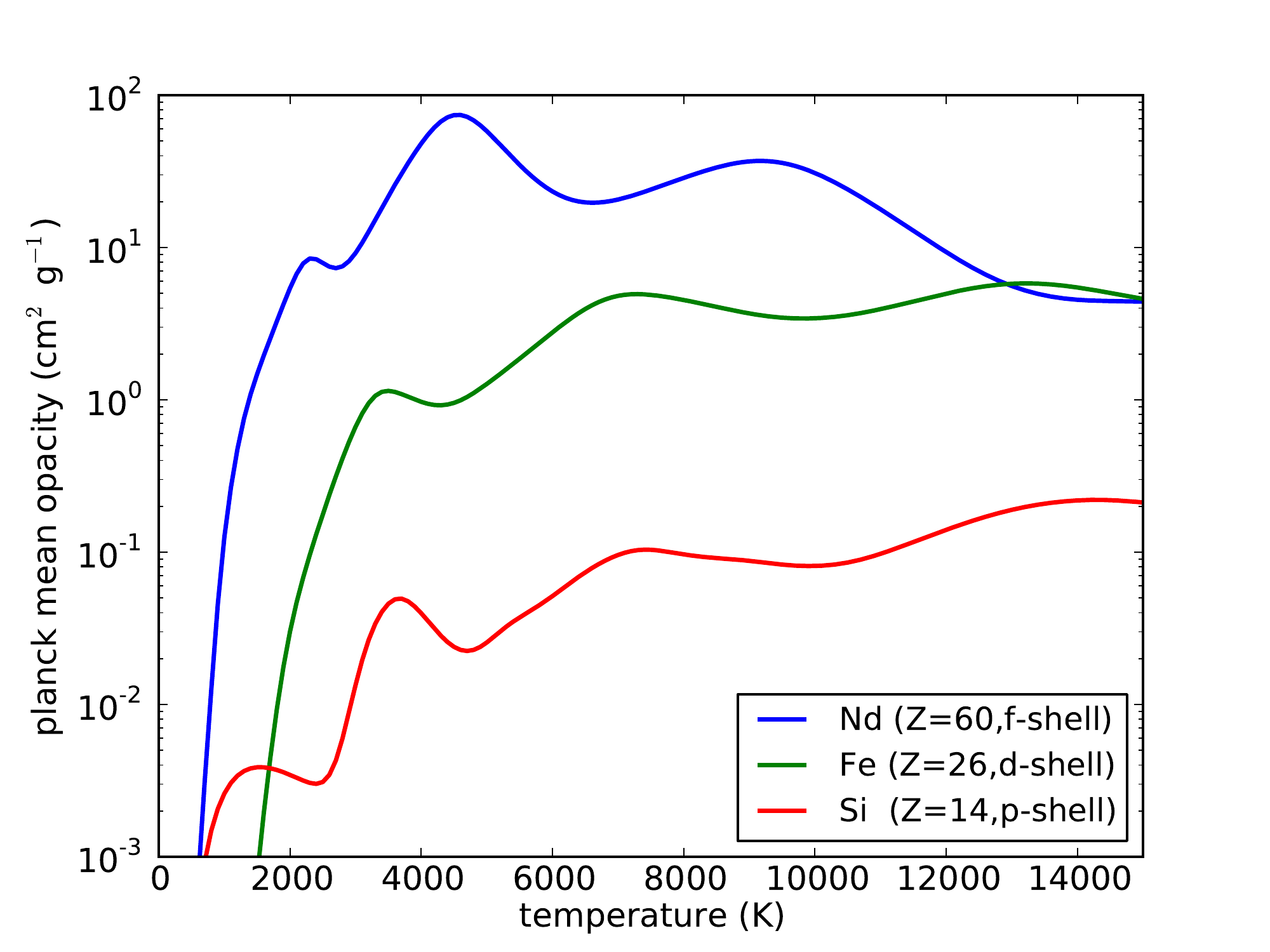}
\caption{ Planck mean expansion opacities for three different
  elements, showing the expected dependence on atomic complexity.  The
  Nd opacities (blue line, $Z=60$, open f-shell) were derived from
  \AS\ models, while the silicon (red line, $Z=14$, open p-shell) and
  iron (green line, $Z=26$, open d-shell) opacities used Kurucz line
  data.  The calculations assume a density $\rho = 10^{-13}~\gcc$ and
  a time since ejection $\texp = 1$~days.
\label{fig:Nd_Fe_Si}}
 \end{figure}

As expected from simple physical arguments, we find that more complex
atoms, in particular the lanthanides, have higher line expansion
opacities. Figure~\ref{fig:Nd_Fe_Si} shows that the Planck mean
opacity of neodymium is a factor $\sim 10-100$ greater than that of
iron, depending on the temperature.  This is roughly consistent with
the estimate one gets by squaring the complexity measure
(equation~\ref{eq:complexity}) to gauge the relative number of strong
lines, ($C_{\rm NdII}/C_{\rm FeII})^2 \approx 22$.

The variation of the mean opacity with temperature
(Figure~\ref{fig:Nd_Fe_Si}) shows several bumps which reflect changes
in the ionization state.  As the temperature increases, the excited
levels become more populated, and the number of optically thick lines
increases.  The opacity therefore increases with temperature until the
gas becomes hot enough to ionize.  This leads to multiple maxima in
the mean opacity curve, each of which occur around the transition
temperatures of the various stages of ionization.  At sufficiently low
temperatures, when the element becomes neutral, the opacities cut off
sharply, and drop exponentially with decreasing temperature due to the
Boltzmann factor in the excited state level populations.

An important property of the lanthanides is that, relative to the iron
group, the opacity remains high at relatively low temperatures.  This
is because the ionization potentials of the lanthanides are generally
$\sim 30\%$ lower than those of the iron group (see Table~1).  For
neodymium, the mean opacity peaks at $T \approx 5000$~K, when the ion
is mostly singly ionized and cuts offs at $T \la 2500$~K when Nd
becomes neutral.  In comparison, the opacity peak for iron occurs at
$T \approx 7000$~K and the neutral cutoff is at $T \la 3500$~K.  The
general persistence of the lanthanide opacity to lower temperatures
has an important impact on the color of the emergent spectra,
contributing to cooler, redder photospheres.

\begin{figure}
\includegraphics[width=3.5in]{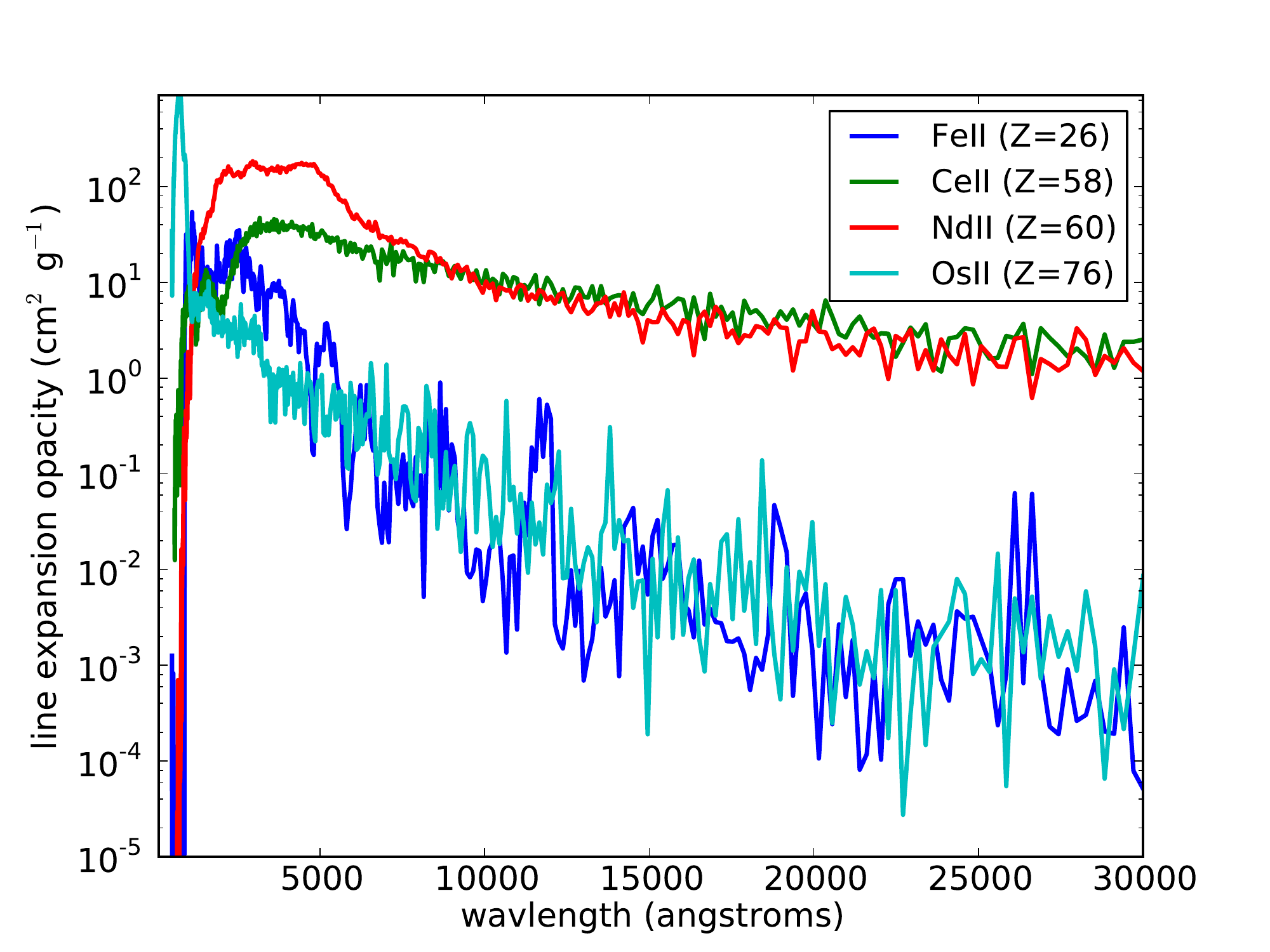}
\caption{ Wavelength dependent line expansion opacities resulting from
  \AS\ derived linelists.  The opacity of the lanthanides (Nd, Ce) is
  much higher then iron and its d-shell homologue, osmium, especially
  in the infrared.
\label{fig:high_Z_exp}}
 \end{figure}

Another important feature of the lanthanide opacity is the wavelength
dependence -- while the opacity decreases to the red (as there are
more lines at bluer wavelengths), the decrease is much slower than
that of the iron group (Figure~\ref{fig:high_Z_exp}).  This is due to
the much denser energy level spacing of the lanthanides, resulting in
a much larger number of $\sim 1$~eV optical/infrared transitions.  The
shallower opacity profile means that the lanthanides can line blanket
not only UV wavelengths, but the entire optical region of the
spectrum. This will influence the color of r-process SNe, as photons
will eventually be re-emitted or fluoresce (through the many lines) to
infrared wavelengths where they may escape more easily.

As seen in Figure~\ref{fig:high_Z_exp}, the opacity of osmium ($Z=76$)
is very similar to that iron, despite the much higher atomic number.
This is not surprising, as osmium is a homologue of iron, with a
nearly half open d-shell.  Similarly, the opacity of the lanthanide
cerium ($Z=58$) is comparable to, though slightly less than, that of
neodymium.  This confirms that species with similar complexity
measures have roughly similar opacities, which we use to derive
approximate opacities for r-process mixtures (\S\ref{sec:rp_mix}).

\subsection{Uncertainties and Comparison to Existing Data}

\begin{figure}
\includegraphics[width=3.5in]{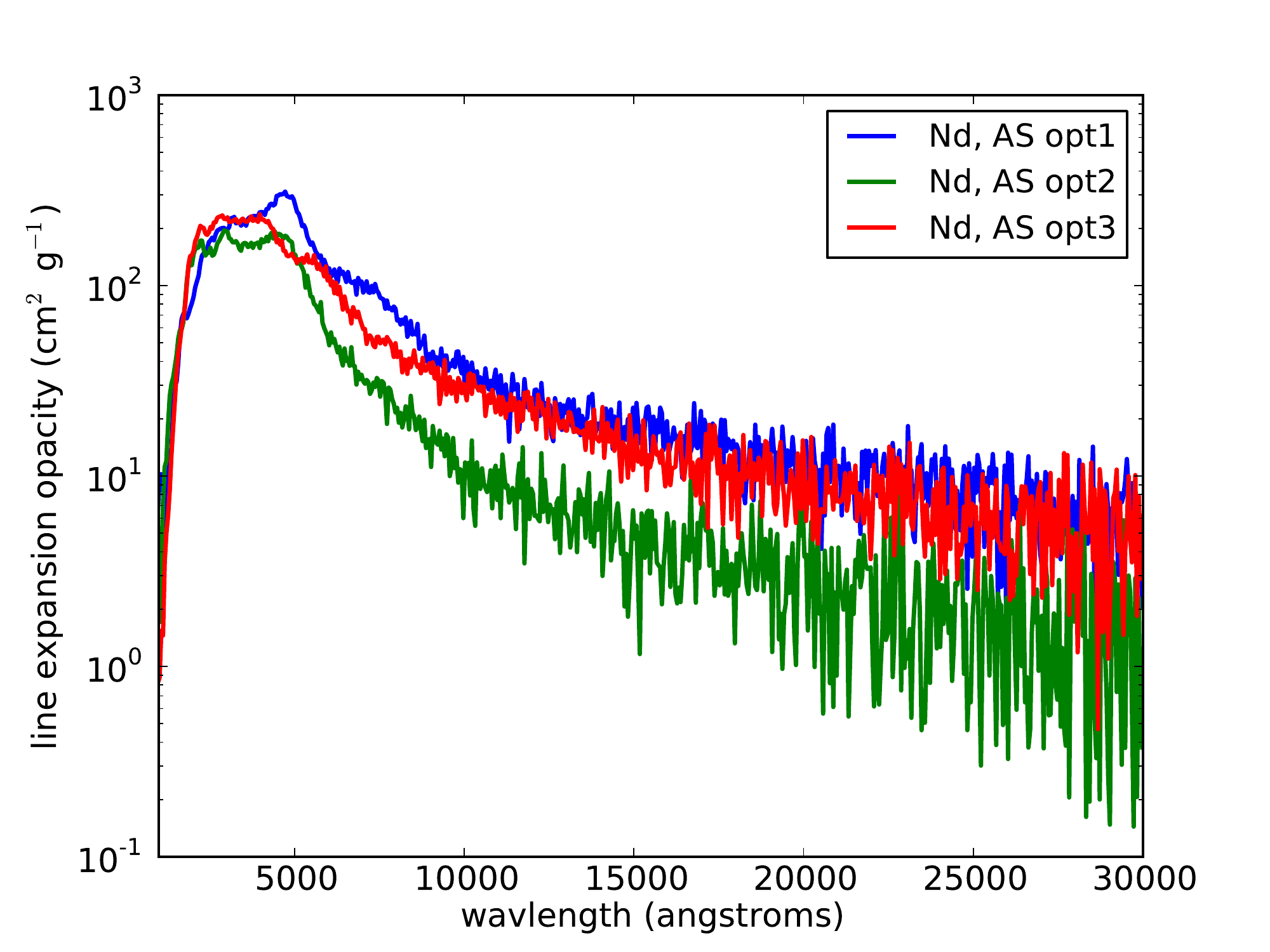}
\caption{ Variations in the wavelength dependent expansion opacity for
  pure neodymium ($Z = 60$) ejecta obtained using different
  \AS\ optimization approaches.  These calculations adopt a density
  $\rho = 10^{-13}$~\gcc, temperature $T = 4000$~K, time since
  ejection $\texp = 1$~days, and a wavelength binning $\Delta \lambda
  = 0.01 \lambda$.
 \label{fig:Nd_vary}}
 \end{figure}
 
Our derived opacities must possess some error, since the \AS\ model
energies do not exactly match the experimental values
(Figure~\ref{fig:Nd_energies}).  To estimate how sensitive the results
are to the detailed level energy structure and configuration ordering,
we examined the NdII opacities derived from the three different
optimization schemes described in \S\ref{sec:AS}.  The resulting
variation provides an estimate of our level of uncertainty.

Figure~\ref{fig:Nd_vary} shows that the opacities calculated using the
{\it opt1} and {\it opt3} models are quite similar, while the {\it
  opt2} model opacities are lower by a factor of $\sim 5$ at some
wavelengths.  The {\it opt2} model has relatively higher energy
levels, and hence smaller excited state LTE level populations, which
is presumably the reason for the lower opacities.  The {\it opt1} and
{\it opt3} models had similar level energies, but the ground state
configuration and ordering were different.  These results suggest that
what matters most to the opacities is the energy level spacing, and
not the exact configuration ordering.  Given that the low lying {\it
  opt3} NdII level energies fairly well reproduce experiment, we
suspect that further fine tuning of the \AS\ model is unlikely to
change the resulting opacity by much more than a factor of $\sim 2$.

\begin{figure*}
\includegraphics[width=3.4in]{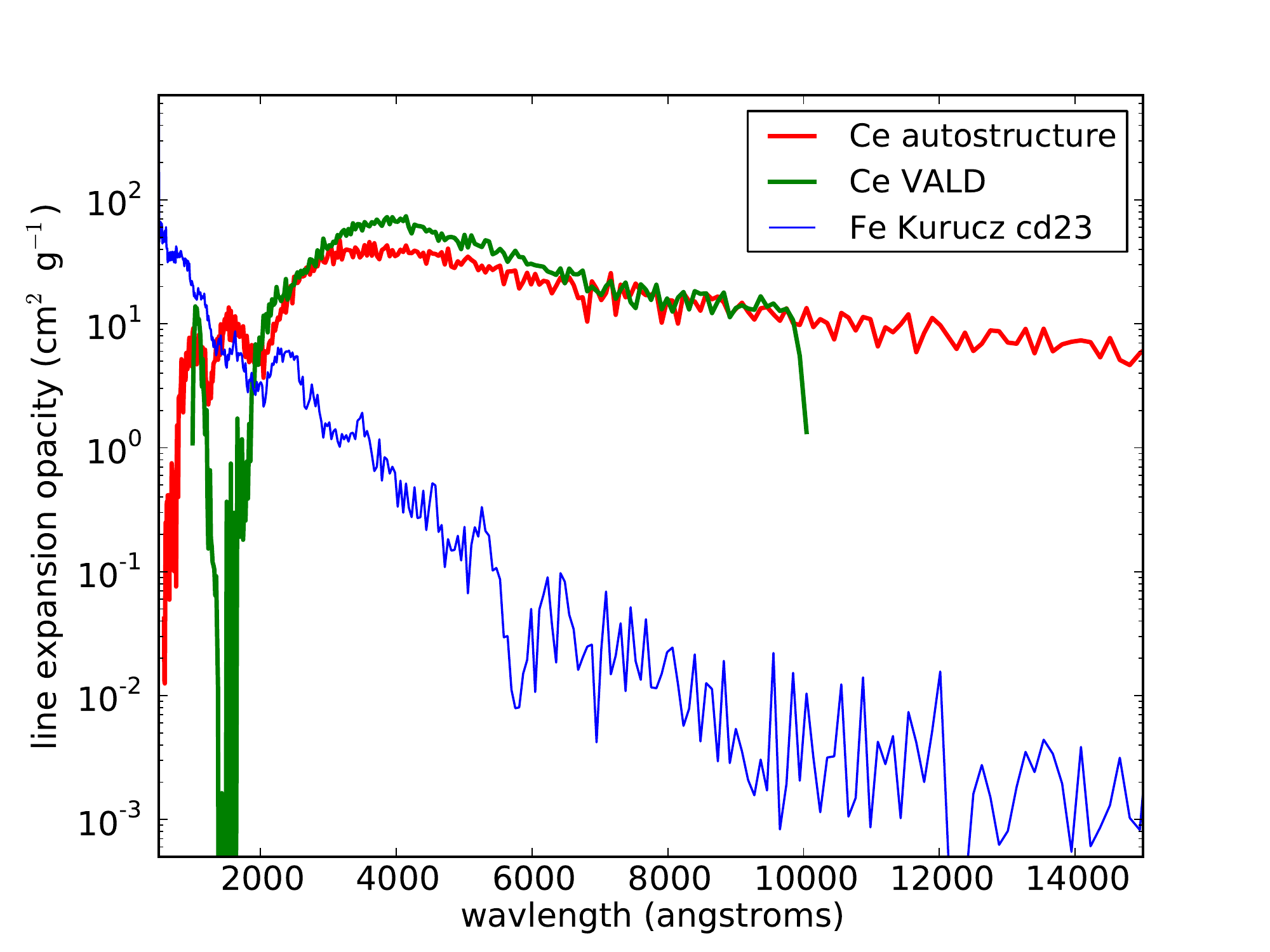}
\includegraphics[width=3.4in]{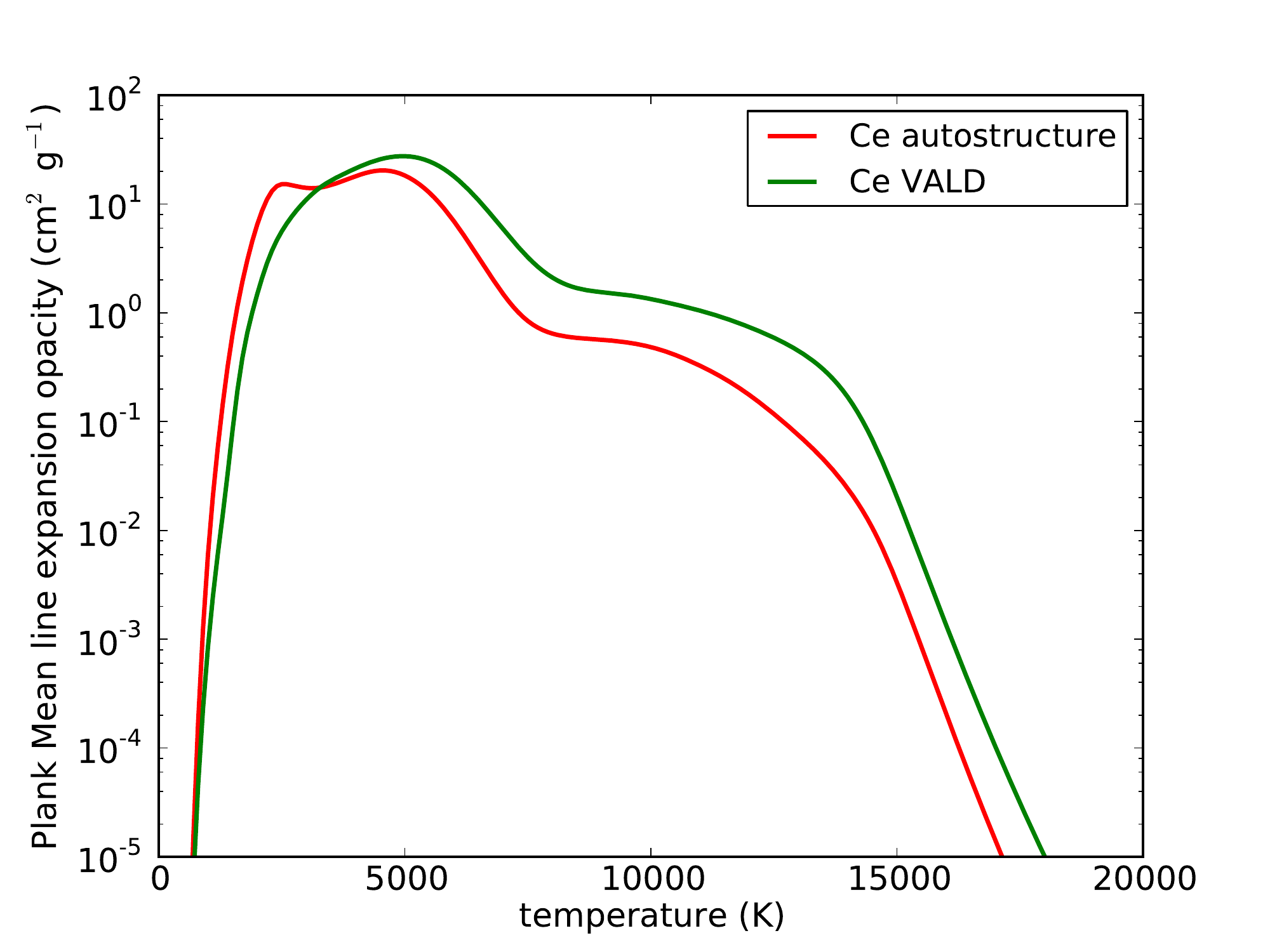}
\caption{ Comparison of the expansion opacity for pure cerium ($Z =
  58$) ejecta, computed using the \AS\ line data (red lines) and the
  VALD linelist (green lines).  The calculations assume a density
  $\rho = 10^{-13}$~\gcc, time since ejection $\texp = 1$~days, and a
  wavelength binning $\Delta \lambda = 0.01 \lambda$.  {\it Left:}
  Line expansion opacity versus wavelength for a temperature $T =
  5000$~K.  At optical wavelengths, the \AS\ results are in reasonably
  good agreement with the VALD; both are orders of magnitude higher
  than the opacity of pure iron ejecta (brown line).  {\it Right:}
  Planck mean opacity as a function of temperature.  Only CeII and
  CeIII are included in the calculation. \label{fig:ce_opacity}}
 \end{figure*}

We have also compared our \AS\ opacities to existing line data from
the VALD database, which collects atomic data from a variety of
sources \citep{VALD_2008}.  The only high-Z ions with enough lines in
VALD to derive expansion opacities are CeII and CeIII, which have
wavelengths and oscillator strengths calculated by the Mons group
\citep{Biemont_1999,Palmeri_2000,Quinet_2004}.  The approach of the
Mons group to atomic structure is the same as that of Kurucz, {\it
  viz.} calculations with Cowan's code utilizing extensive
experimental energies.  In Figure~\ref{fig:ce_opacity}, we compare the
expansion opacities of Ce calculated using the VALD linelist and our
own \AS\ list.  The agreement in both the mean and wavelength
dependent values is good to a factor of $\sim 2$. Our conclusions
about the size and wavelength dependence of the lanthanide opacities
are therefore confirmed when using radiative data from independent
structure calculations.

\section{Opacities of r-process mixtures}
\label{sec:rp_mix}

\begin{figure}
\includegraphics[width=3.50in]{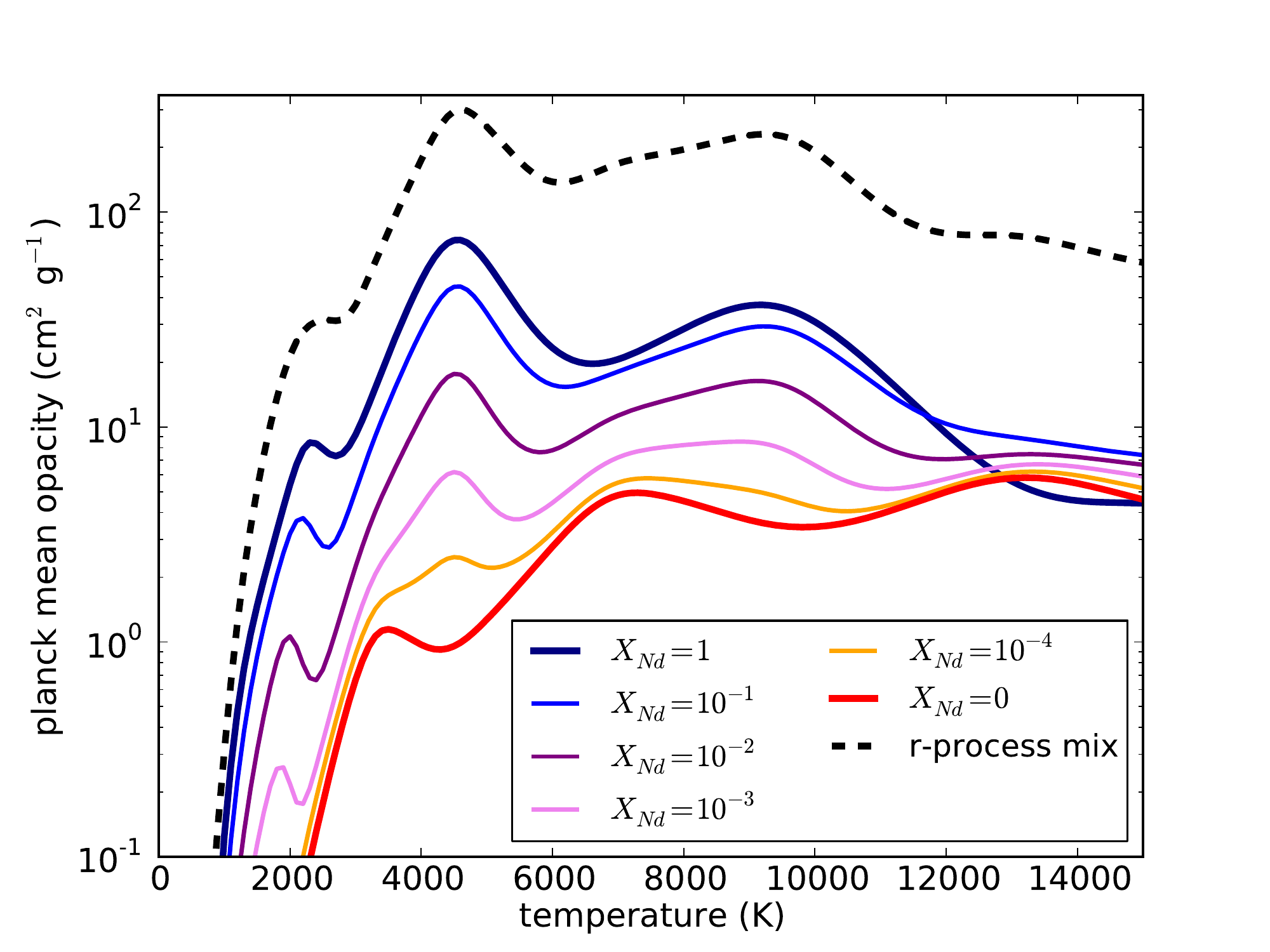}
\caption{Dependence of the mean expansion opacity on the abundance of
  lanthanides.  The solid lines show the Planck mean opacity for
  various mass fractions of neodymium in a mixture with iron. The
  dashed line shows the opacity of the approximate r-process mixture
  (with all 14 lanthanides) discussed in \S\ref{sec:rp_mix}.
 \label{fig:Nd_abun}}
 \end{figure}

Although we have only calculated atomic structure models for a few
ions, the results (Figure~\ref{fig:high_Z_exp}) suggest that ions of
similar complexity have roughly similar opacities.  This allows us to
construct approximate r-process mixtures based on the representative
cases.

In an r-process mixture, the abundance of any individual lanthanide is
relatively low $(\la 1\%)$.  Nevertheless, these species likely
dominate the total opacity.  In fact, the opacity will depend rather
weakly on the exact lanthanide abundance.  This is because for the
conditions found in NSM ejecta, many of the strong lanthanides lines
are extremely optically thick ($\tau_s \gg 1$).  Such lines contribute
equally to the expansion opacity regardless of the ion's abundance,
just as long as that abundance remains high enough to keep $\tau_s$
above unity.

We illustrate this weak dependence on lanthanide abundance in
Figure~\ref{fig:Nd_abun}, by computing the opacity of a mixture of
neodymium and iron.  Decreasing the Nd mass fraction by a factor of
ten (from 100\% to 10\%) only reduces the total opacity of the mixture
by $\sim 40\%$. Decreasing the Nd mass fraction by two orders of
magnitudes (from 100\% to 1\%) reduces the total opacity of the
mixture by a factor of 5.  We find that the Nd opacity dominates over
that of iron as long as its mass fraction is $\ga 10^{-4}$.

The actual r-process ejecta from NSMs will be a heterogeneous mixture
of many high $Z$ elements.  This multiplicity of species should
enhance the opacity, as each ion contributes a distinct series of
lines.  To estimate the opacity of the mixture we assume the line data
of Nd is representative of all f-shell species (the lanthanides) and
that iron is representative of all d-shell elements.  We ignore the
s-shell and p-shell elements since their opacities will be very low.
We then construct the expansion opacity of the mixture by generalizing
eq.~\ref{eq:ex_opacity}
\begin{equation}
\kappa_{\rm mix}(\lambda) = 
\sum_Z
\frac{\xi_Z}{\rho c \texp}  \sum_i \frac{\lambda_i}{\dls } 
 \biggl( 1 - \exp[-\tau_i(\rho_Z)] \biggr)
\end{equation}
where the sum $Z$ runs over the representative ions (in this case only
Fe and Nd) and the quantity $\xi_Z$ specifies the total number of
elements represented by each.  For Nd, $\xi_Z = 14$ to account for all
14 lanthanides, while for iron $\xi_Z = 30$ to account for all d-shell
elements between $21 \le Z \le 80$.  The quantity $\rho_Z = X_Z \rho$
is the density of the representative elements, where $X_Z$ is the mass
fraction of each.  As an illustrative r-process mixture, we assume
that the average mass fraction of each lanthanide is $X_f = 1\%$, and
the average d-shell species fraction is $X_d = 2\%$.  The remainder of
the composition was taken to be calcium (s-shell) as a neutral filler.
We used the Nd line data from the {\it opt3} structure model, and the
iron line data from the Kurucz CD23 list.

The dashed line in Figure~\ref{fig:Nd_abun} shows the Planck mean
opacity of our approximated r-process mixture.  Because each of the 14
lanthanides is assumed to contribute independently in the sum, the
total opacity is essentially 14 times that of the mixture with only
1\% Nd.  At certain temperatures when the lanthanide opacity dips, the
d-shell opacity makes a comparable contribution.  We note that opacity
of the mixture can approach the saturation level discussed in
\S\ref{sec:sobolev}, such that our assumption that the strong lines do
not overlap can be called into question.

\section{Spectra of NS merger ejecta}
\label{sec:spectra}

 To illustrate the general effect of our r-process opacities on the
emission from NSM ejecta, we have calculated model spectra using the
\Sedona\ radiation transport code \citep{Kasen_MC}.  A more
comprehensive discussion of the light curves and colors of these
transients, and their dependence on the ejecta parameters, is given in
\cite{Barnes_2013}.

As a simple, fiducial ejecta model, we considered a spherically
symmetric, homologously expanding remnant with a broken power-law
density profile.  The total ejecta mass was taken to be $M_{\rm ej} =
0.01~\msun$ and the kinetic energy $E = 1/2 M_{\rm ej} v_c^2$ with a
characteristic velocity $v_c = 0.1c$.  The transport calculations
assume the ionization/excitation state is given by LTE, and that the
line source function is described by the Planck function, i.e., the
medium is purely absorbing.  In reality, the probability of absorption
in lines may be small, with fluorescence being the more likely result
of line interactions.  However, supernova transport calculations have
shown that for complex ions, repeated fluorescence among a multitude
of lines has an approximately thermal character \citep{Pinto_2000a,
  Kasen_IR}.  We use the opacities of an r-process mixture derived in
\S\ref{sec:rp_mix}.

\begin{figure}
\includegraphics[width=3.5in]{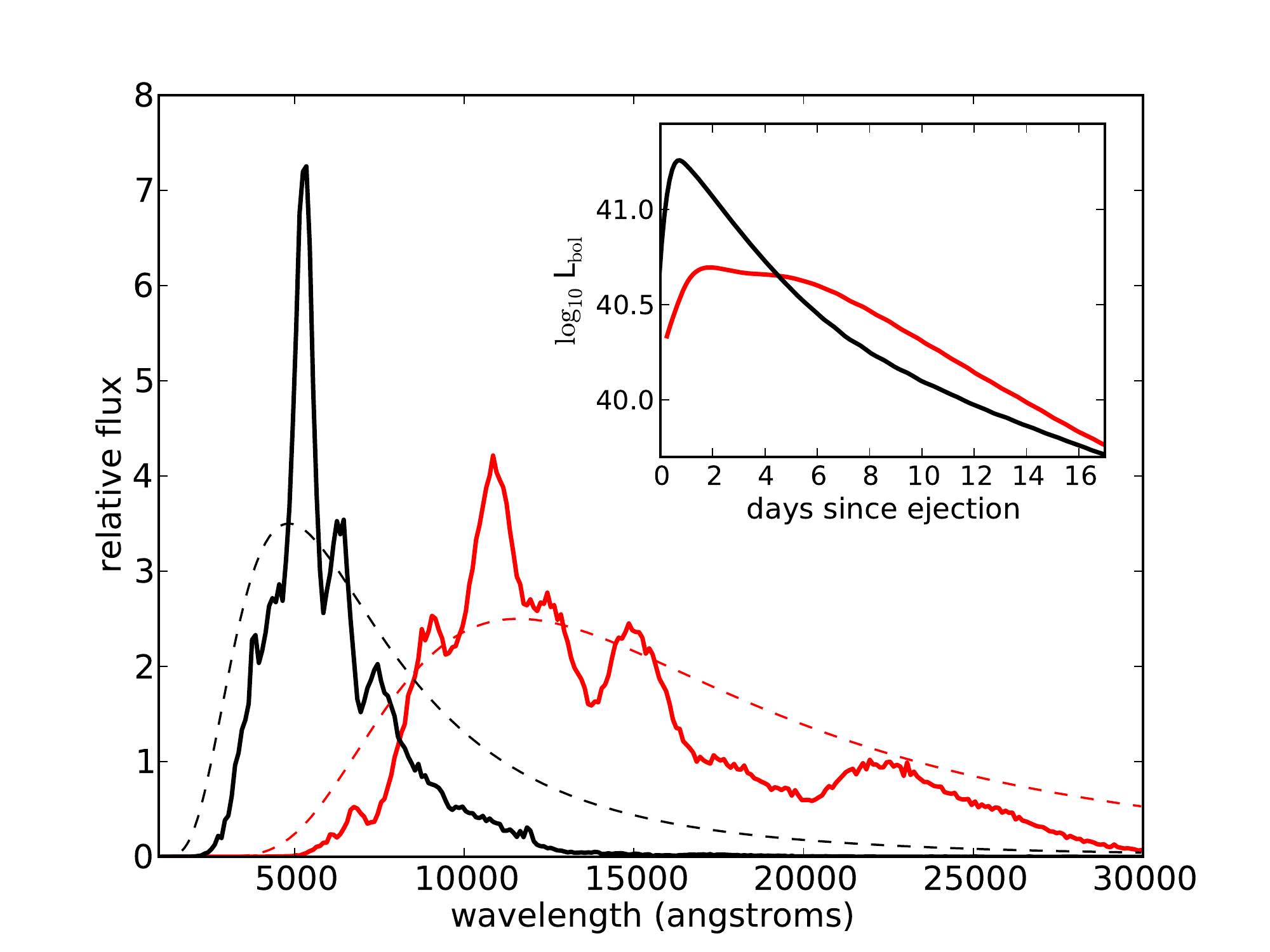}
\caption{Synthetic spectra (2.5 days after mass ejection) of the
  r-process SN model described in the text, calculated using either
  Kurucz iron group opacities (black line) or our \AS\ derived
  r-process opacities (red line).  For comparison, we overplot
  blackbody curves of temperature $T = 6000$~K (black dashed) and $T =
  2500$~K (red dashed).  The inset shows the corresponding bolometric
  light curves assuming iron (black) or r-process (red) opacities.
 \label{fig:rp_spectra}}
 \end{figure}

 \begin{figure}
\includegraphics[width=3.5in]{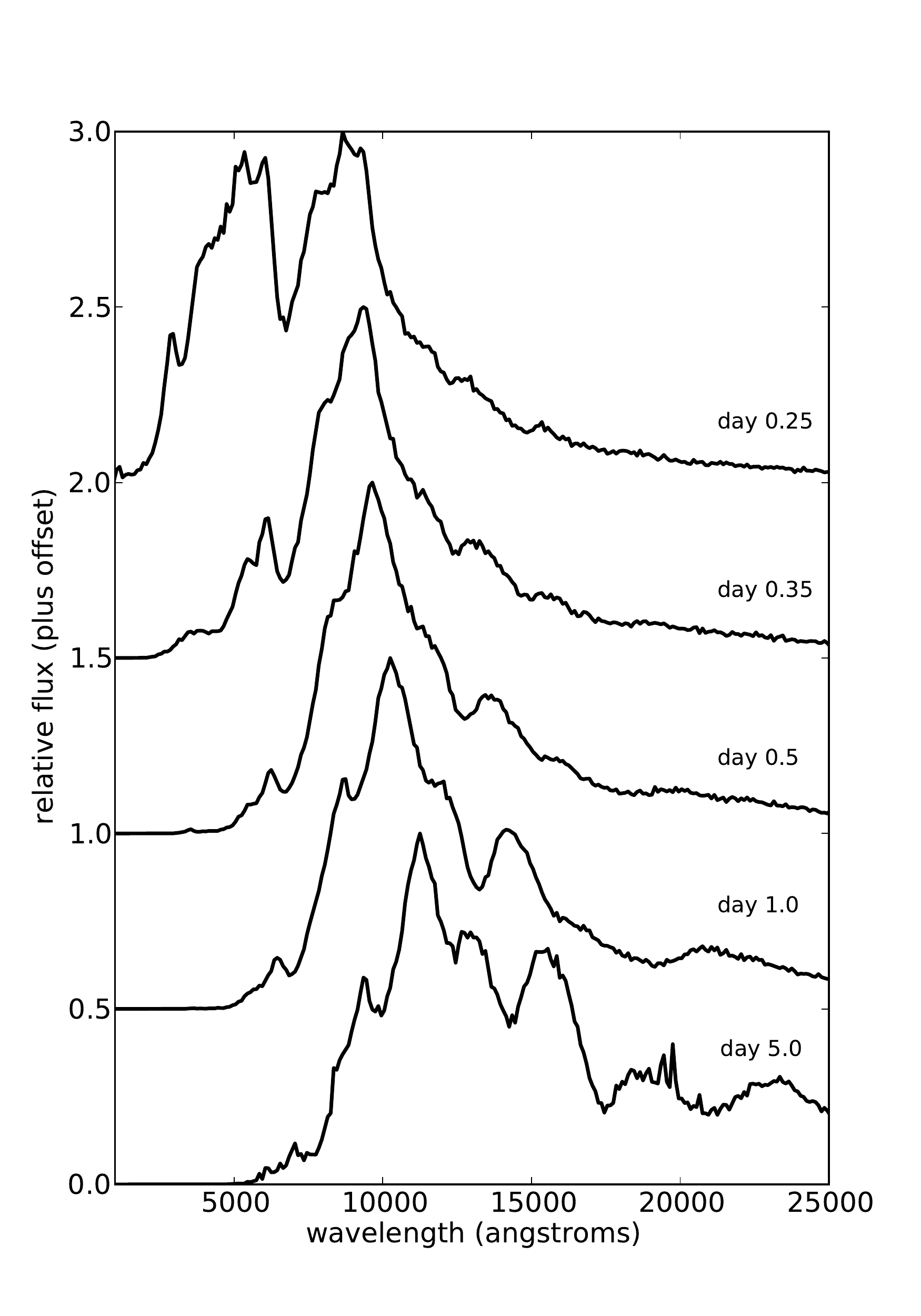}
\caption{Synthetic spectra time series of the r-process SN model
  described in the text.  The times since mass ejection are marked on
  the figure.
 \label{fig:spec_series}}
 \end{figure}

The high opacity of r-process material has a significant impact on the
predicted radioactive transients from NSMs
(Figure~\ref{fig:rp_spectra}).  Compared to previous calculations
(which assumed iron-like opacities) the predicted bolometric light
curve is of much longer duration, $\sim 1$ week as opposed to $\sim 1$
day.  This is due to the longer effective diffusion time through the
more opaque ejecta.  The peak luminosity is also lower, as the
radiation suffers greater loses due to expansion over this period.
The spectrum at 2.5 days after merger is much redder, with most of the
flux emitted in the near infrared ($\sim 1~\mu$m).  Due to the extreme
line blanketing at bluer wavelengths, the photons are eventually
redistributed (through lines) to the infrared, where the opacities are
lower and radiation can escape more readily.

Other than the unusually red color, the r-process spectra generally
resemble those of ordinary SNe, and in particular those with high
expansion velocities (e.g., the hyper-energetic Type~Ic event,
SN~1998bw \citep{Galama_1998}).  The continuum flux, which is produced
by emission in the Doppler-broadened forest of lines, resembles a
blackbody with a few broad ($\sim 200$~\AA) spectral features.  It is
not easy to associate these features with either absorption or
emission from a single line; instead they arise from blends of many
lines.  Because our atomic structure models do not accurately predict
line wavelengths (and we only include lines of Nd and Fe), the
location of the features in our synthetic spectra are not to be
trusted.  Nevertheless, the general appearance of the model spectra are likely qualitatively
correct. One can anticipate where features are most likely to appear
by examining the energy spacing of the low lying levels of the
lanthanides.

Figure~\ref{fig:spec_series} shows the time evolution of the synthetic
spectra.  At the earliest times ($\lesssim 0.25$~days after ejection)
some flux emerges at optical wavelengths, but this phase is short
lived.  By day $0.5$, the optical emission has faded, and the spectra
evolve relatively slowly thereafter, with effective blackbody
temperatures steady in the range $T \approx 2000-3000$~K.  The
temporal evolution can be understood by considering the mean opacity
curves (e.g., Figure~\ref{fig:Nd_Fe_Si}).  At early times, the ejecta
is relatively hot ($\gtrsim 4000$~K) throughout, and the opacity is
roughly constant with radius.  By day $\sim 0.3$, however, the
outermost layers have cooled below $\lesssim 3000$~K, and the
r-process opacities drop sharply due to lanthanide recombination.  The
ejecta photosphere forms near the recombination front (as overlying
neutral layers are essentially transparent) which regulates the
effective temperature to be near the recombination temperature. This
behavior is similar to the plateau phase of the (hydrogen rich)
Type~IIP SNe, although in this case the opacity is due to line
blanketing, not electron-scattering.  More importantly, the
temperature at the recombination front ($T_{I} \sim 2500$~K) is a
factor of $\sim 2$ lower for r-process ejecta, as the ionization
potentials of the lanthanides ($\sim 6$~eV) are lower than that of
hydrogen ($\sim 13.6$~eV).  It is possible that non-thermal
ionization from radioactive decay products 
(see \S\ref{sec:LTE}) become important at these phases and prevent 
the outer layers from ever becoming 
completely  neutral.

 \begin{figure}
\includegraphics[width=3.5in]{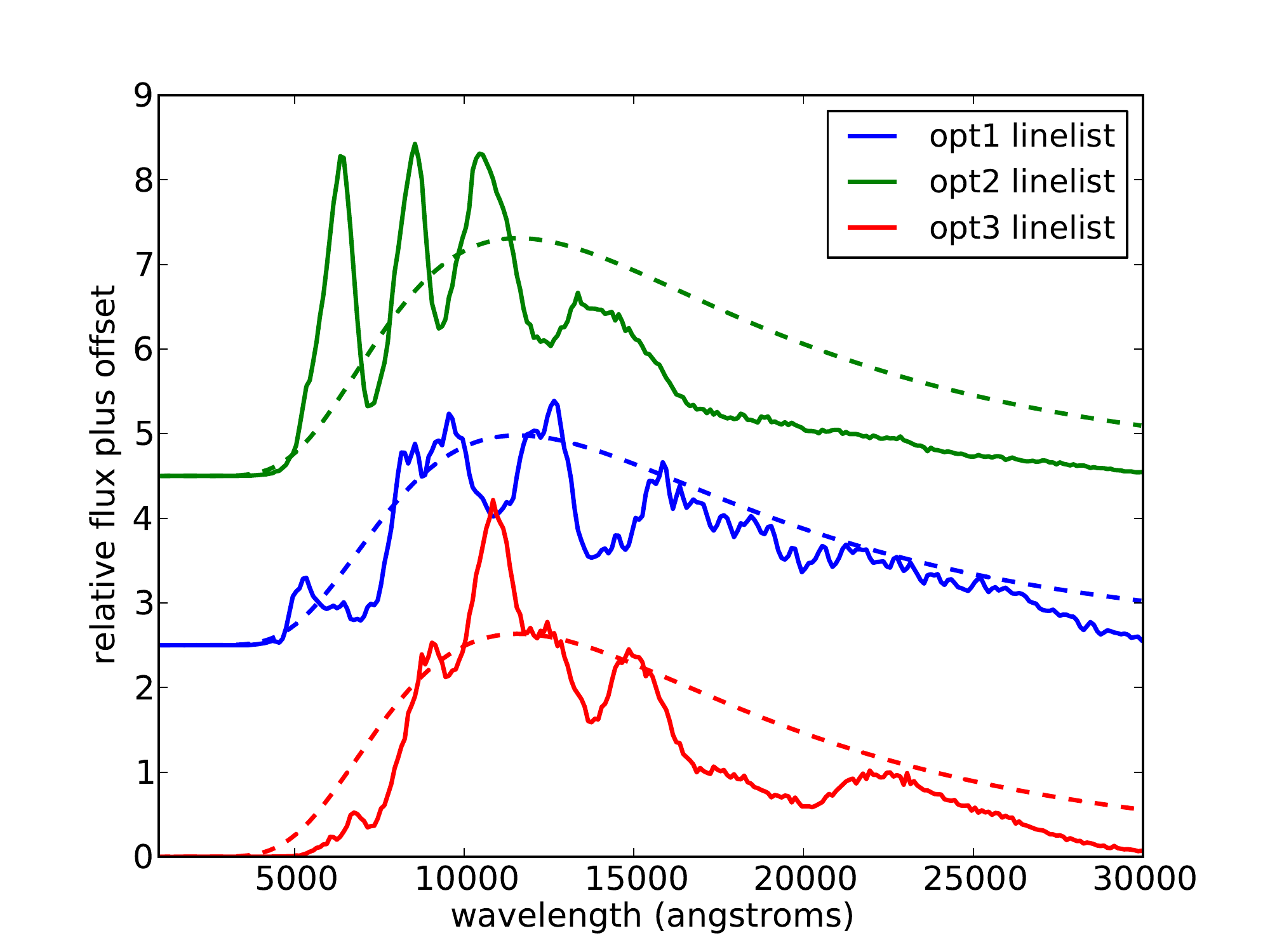}
\caption{Effect of varying the atomic structure model on the predicted
  observables of an r-process SN. The figure shows the synthetic
  spectrum (2.5 days after ejection) calculated using \AS\ linelists
  under different optimization schemes.  The dashed lines show, for
  comparison, blackbody curves of temperature $T = 2500$~K.
 \label{fig:comp_spec}}
 \end{figure}
 
Our calculated SED's are somewhat sensitive to the atomic structure
model used to generate the line data.  Figure~\ref{fig:comp_spec}
compares calculations using line data from the different
\AS\ optimization runs ({\it opt1, opt2,} and {\it opt3}).  The
observed differences can be taken as some measure of the uncertainty
resulting from inaccuracies in our atomic structure calculations.
Notably, the spectrum calculated using the {\it opt2} linelist has
significantly higher flux in the optical ($\sim 6000$~\AA).  This is
presumably due to the lower overall opacity of the {\it opt 2} model
(Figure~\ref{fig:Nd_vary}). Given the superior match of the {\it opt3}
model to the experimental level data, we consider the spectral
predictions using this line data to be the most realistic, however it
is clear that some significant uncertainties remain.

Another concern for the spectrum predictions is the potential
breakdown of the Sobolev approximation (see \S\ref{sec:sobolev}).  At bluer wavelengths, the
mean spacing of strong lines can become less than the intrinsic
(presumed thermal) width of the lines, which violates the assumptions
used to derive an expansion opacity.  It is not immediately clear how
this will impact the results.  Two lines that overlap exactly will
behave like a single line, which suggests that an overlapping line
should be discounted, not double counted.  In this case, the opacity
should saturate at a maximum value $\kappa_{\rm sat}$
(equation~\ref{eq:kappa_sat}). On the other hand, in this saturation
limit, photons can no longer escape lines by redshifting past them, as
there are no longer any optically thin ``windows" between lines.  The
individual line optical depths then become relevant, and the effective
opacity may be larger than one would estimate from the expansion
formalism.  In practice, the impact of line overlap may not be so
dramatic -- the opacity generically declines to longer wavelength, and
the net effect of the transport is to distribute photons to the
red/infrared where saturation may no longer be an issue.  To fully address
the question will require  transport calculations
that dispense with the Sobolev approximation and resolve individual
line profiles.

\section{Summary and Conclusions}

The opacity of r-process ejecta is orders of magnitude higher than
that of ordinary supernova debris, a fact we have demonstrated using
new atomic structure calculations (Figure~\ref{fig:high_Z_exp}),
pre-existing line data (Figure~\ref{fig:ce_opacity}), and simple
physical counting arguments (Figure~\ref{fig:complexity}).  There are
two physical reasons for the high opacity: 1) {\it Complexity:} The
r-process composition includes rare elements with complex valence
electron structure, in particular the lanthanides which have an open
valence $4f$-shell.  Such elements have a significantly greater number
of levels and lines, which results in an overall higher expansion
opacity; 2) {\it Multiplicity:} The r-process produces a heterogenous
mixture of many elements, each of which contributes a distinct series
of lines.  Since the expansion opacity depends on the total number of
strong lines (and not the strength of any individual line) this
diversity of the mixture enhances the opacity relative to a more
homogenous composition.

While our r-process opacity calculations offer a significant
improvement over previous estimates (which were little more than
educated guesses) several uncertainties remain.  The \AS\ models only
approximate the level structure of the high-Z elements, and so do not
correctly predict the wavelength of individual transitions.
Fortunately, the pseudo-continuum opacities depend only on the
statistical distribution of lines and are fairly robust, although our
numerical experiments indicate uncertainties at the factor of $\sim 2$
level.  In the future, we can iteratively tune the structure models to
better reproduce the observed energy levels, although this is a time
consuming process.  Moreover, many of the high-Z ions lack good
experimental level data.

Another, more important, uncertainty is that we have used the
radiative data for one species (Nd, $Z=60$) to represent all
lanthanides.  In fact, not all lanthanides are equal -- the ions whose
valence f-shell is nearly open (e.g., La, $Z=56$) or closed (Yb,
$Z=70$) are considerable less complex, and should have correspondingly
lower opacities.  On the other hand, gadolinium (Gd, $Z=64$) has a
nearly half filled f-shell, and is one of the most complex species on
the periodic table.  Simple counting arguments
(Figure~\ref{fig:complexity}) suggest that Gd may have an opacity
$\gtrsim 10$ times that of Nd.  We therefore suspect that our current
opacities {\it underestimate} the true values for an r-process mixture
where Gd is present at the $\sim 1\%$ level.  In future work, we will
quantify the line data for all lanthanides.  The actinides should also
be considered, as they are open f-shell as well, although only a few
species (e.g., uranium) likely have high enough abundance to make a
difference.  In total, the atomic data required for r-process
opacities is massive, involving numerous structure calculations and
many billions of lines.

The high opacity of r-process material has significant implications
for discovering and interpreting the radioactively powered transients
associated with NSMs.  With more realistic opacities, the predicted
light curves are of longer duration and dimmer at peak
\citep[see][]{Barnes_2013}.  Perhaps even more important to detection
is that the SED is shifted into the infrared, peaking at $\sim
1~\mu$m.  This is due to the strong line blanketing at optical
wavelengths, which pushes the photosphere out to cooler layers ($T
\sim 2500$~K) where the lanthanides recombine and become more
transparent.  Other than the very red color, the spectra qualitatively
resemble those of other high-velocity SNe, with a pseudo blackbody
continuum and broad ($\sim 200$~\AA) spectral features.

The predicted emission at optical wavelengths is somewhat sensitive to
the details of the opacity and its associated uncertainties.  Spectra
calculated using our {\it opt2} radiative data were fairly bright in
the V-band ($\sim 5000$~\AA), while those calculated using {\it opt3}
data had almost no flux at these wavelengths. Because the level
structure of our {\it opt3} model agrees better with experiment (and
given that the overall opacity may be even higher than our Nd-based
estimates) we consider the latter case to be the more likely reality.
However, the opacities are still not fully converged, and we cannot
altogether rule out the possibility that r-process SNe may emit some
persistent emission in the optical.

These results suggest that (to the extent possible) it is worthwhile
to search for and/or follow up gravitational wave sources at red or
infrared wavelengths.  Optical surveys, however, will still be 
sensitive to a radioactive transient if some component
of the ejecta is nearly lanthanide free.
In fact, it is likely that, in addition to the tidal tails, a second
component of lighter elements ($Z \lesssim 50$) is ejected from a
post-merger disk wind.  If this wind includes radioisotopes with
appropriate half-lives (e.g., \Nifs), the light curve may be
relatively bright and peak in the optical \citep{Barnes_2013}.  A detailed understanding
of the wind nucleosynthesis and mixing with the tails is important, as
our results suggest that contamination by lanthanides at just the
$10^{-3}$ level may significantly raise the opacities.

Beyond detection, a significant observational challenge will be
confirmation that a transient is indeed due to a NSM.  There are
likely many classes of stellar explosions that produce low mass
ejections of radioactive material \citep[e.g.,][]{Bildsten_2007,
  Moriya_2010}, and an increasing number of fast, faint transients
have been observed at optical wavelengths
\citep[e.g.,][]{Kasliwal_2010,Perets_2010,Kasliwal_2012,Foley_2012}.
Fortunately, our calculations demonstrate that the optical properties
of r-process ejecta differ dramatically from transients due to lower
mass isotopes.  A key distinguishing features is that the r-process
SED peaks in the infrared with a nearly constant color temperature
regulated to the lanthanide recombination temperature, $\sim 2500$~K.
The population of brief, infrared variables is mostly unknown, but it
is possible that the infrared transient sky is much cleaner than the
optical one.
 
By comparing observations of an r-process SN to light curve models,
one could presumably constrain the mass of heavy nuclei ejected in an
compact object merger, which would go a long way to understanding the
unknown site(s) of the r-process.  One would like to go further and
spectroscopically study the abundance distribution of the outflows.
That will be challenging -- the lines are heavily blended and we do
not yet have accurate wavelengths for most of them.  In the future,
though, we can refine the line data by tuning structure models to
match experimental data (where available) and can use radiative transport
calculations to quantify how global abundances variations affect the
blended features. While measuring a detailed abundance pattern will be
hard, spectroscopic modeling should permit strong constraints
on the amount and gross composition of ejecta.

This assumes that radioactive r-process transients exist and that we
can find them. Admittedly, we test dangerous waters any time that,
lacking observational input, we attempt to describe a new
astrophysical phenomenon on purely theoretical grounds.  The situation
here is a step more treacherous; not only must we rely on simulations
of a complex macroscopic system, even the {\it microscopic} structures
of our ions are model-based.  Obviously, observational input is needed
to ground the theory.  In the meantime, the numerical experiment
presented in \S\ref{sec:sn1a} may offer a bit of comfort.  In that
example, we calculated the light curve of an obliviously crude model
of a SN~Ia (a uniform \Nifs\ blob) using opacities derived entirely
from {\it ab-initio} atomic structure models.  Despite deliberately
ignoring decades of work in the field, our ``supernova from scratch"
predictions weren't all that bad, and certainly good enough to allow
one to search for and identify a thermonuclear event.  There is then
reason to hope that the predictions for r-process SNe are (or will
soon be) good enough for us to know one when we see one.

\begin{acknowledgements}
This research has been supported by a Department of Energy Office of Nuclear
Physics Early Career Award, and by the Director, Office of Energy
Research, Office of High Energy and Nuclear Physics, Divisions of
Nuclear Physics, of the U.S. Department of Energy under Contract No.
DE-AC02-05CH11231.  This work is supported 
in part by an NSF Division of Astronomical Sciences collaborative 
research grant AST-1206097.
The work of NRB was supported by STFC
(ST/J000892/1).  We are grateful for computing time made available
the National Energy Research Scientific Computing Center, which is supported by the Office of Science of the U.S. Department of Energy under Contract No. DE-AC02-05CH11231. 
\end{acknowledgements}

\end{document}